\documentclass{article}

\usepackage{arxiv}

\usepackage{natbib}
\usepackage[utf8]{inputenc} % allow utf-8 input
\usepackage[T1]{fontenc}    % use 8-bit T1 fonts
\usepackage[hidelinks]{hyperref}
\usepackage{url}            % simple URL typesetting
\usepackage{booktabs}       % professional-quality tables
\usepackage{amsfonts}       % blackboard math symbols
\usepackage{nicefrac}       % compact symbols for 1/2, etc.
\usepackage{microtype}      % microtypography
\usepackage{lipsum}
\usepackage{graphicx}
\graphicspath{ {./images/} }

\usepackage[section]{algorithm}
\usepackage[noend]{algpseudocode}
\usepackage{bm}
\usepackage{caption}
\usepackage{contour}
\usepackage[acronym]{glossaries}
\usepackage{multirow}
\usepackage{physics}
\usepackage{subcaption}
\usepackage{tabularx}
\usepackage{tikz}
\usepackage{tikzfill}
\usepackage{xspace}

\usepackage{pgfplots}

\usepackage[nameinlink,noabbrev]{cleveref}

\newacronym{mc}{MC}{Monte Carlo}
\newacronym{mcmc}{MCMC}{Markov chain Monte Carlo}

\newacronym{mh}{MH}{Metropolis-Hastings}
\newacronym{mala}{MALA}{Metropolis-adjusted Langevin algorithm}
\newacronym{hmc}{H²MC}{Hamiltonian Monte Carlo}

\newacronym{pt}{PT}{path tracing}
\newacronym{bdpt}{BDPT}{bidirectional path tracing}
\newacronym{jrlt}{JRLT}{Jump Restore Light Transport}
\newacronym{rb}{RB}{Rao--Blackwellization}

\newacronym{mae}{MAE}{mean absolute error}
\newacronym{mse}{MSE}{mean squared error}
\newacronym{mrse}{MRSE}{mean relative squared error}
\newacronym{mape}{MAPE}{mean absolute percentage error}

\newacronym{sde}{SDE}{stochastic differential equation}

\newacronym{fps}{FPS}{frames per second}
\newacronym{spp}{SPP}{samples per pixel}

\let\originalleft\left
\let\originalright\right
\renewcommand{\left}{\mathopen{}\mathclose\bgroup\originalleft}
\renewcommand{\right}{\aftergroup\egroup\originalright}

\newcommand\declaresymbol[2]{\newcommand{#1}{\TextOrMath{$#2$\xspace}{#2}}}

\declaresymbol\exponentialparameter\alpha

\declaresymbol\normalvariable\xi

\declaresymbol\timedomain T

\declaresymbol\prevtimepoint s
\declaresymbol\timepoint t
\declaresymbol\point x
\declaresymbol\pointset B

\declaresymbol\targetdistribution\pi
\declaresymbol\targetdensity p
\declaresymbol\referencemeasure\lambda

\declaresymbol\integrand f
\declaresymbol\secondintegrand g

\declaresymbol\ergodicaverage S

\declaresymbol\process X
\declaresymbol\localprocess Y

\declaresymbol\localsemigroup\kappa

\declaresymbol\generator A
\declaresymbol\localgenerator L
\declaresymbol\globalgenerator G

\declaresymbol\regenerationdistribution\mu
\declaresymbol\regenerationdensity u

\declaresymbol\killingrate k
\declaresymbol\killingrateconstant{k_0}

\declaresymbol\lifetime\tau

\declaresymbol\markovchain M
\declaresymbol\transitionkernel\kappa

\declaresymbol\holdingrate h
\declaresymbol\combinedrate c

\declaresymbol\operator R

\declaresymbol\wienerprocess W

\declaresymbol\drift b
\declaresymbol\diffusioncoefficient\sigma

\declaresymbol\covariance\Sigma
\declaresymbol\rotationscale\gamma
\declaresymbol\antisymmetry U

\declaresymbol\correctionoperator C
\declaresymbol\testfunction\varphi

\declaresymbol\dimensionindex i
\declaresymbol\dimension d
\declaresymbol\reduceddimension k

\makeatletter
\newcommand{\setword}[2]{%
    \phantomsection
    #1\def\@currentlabel{\unexpanded{#1}}\label{#2}%
}
\makeatother

\algrenewcommand\algorithmicrequire{\textbf{Input:}}
\algrenewcommand\algorithmicensure{\textbf{Output:}}

\algdef{SE}[SUBALG]{Indent}{EndIndent}{}{\algorithmicend\ }%
\algtext*{Indent}
\algtext*{EndIndent}

\definecolor{DarkGreen}{rgb}{0, .6, 0}
\newcommand{\AlgCommentTemplate}[2]{\hfill{\fontsize{7}{6}\selectfont\textcolor{DarkGreen}{\text{#1\;#2}}}}

\newcommand{\AlgCommentLeft}[1]{\AlgCommentTemplate{$\leftarrow$}{#1}}

\definecolor{myblue}{RGB}{0, 114, 178}
\definecolor{myorange}{RGB}{230, 159, 0}
\definecolor{myteal}{RGB}{0, 158, 115}
\definecolor{mypurple}{RGB}{204, 121, 167}


\newtheorem{definition}{Definition}

\title{Diffusion Restore:\\Real-Time Markov Chain Monte Carlo Light Transport}

\author{
  Sascha Holl\thanks{Corresponding author.}\\
  Max Planck Institute for Informatics\\
  Saarland University\\
  Saarbrücken, Germany\\
  \texttt{sascha.holl@gmail.com}\\
  \And
  Gurprit Singh\\
  Advanced Micro Devices, Inc. (AMD)\\
  Munich, Germany\\
  \texttt{gurprit.singh@amd.com}\\
  \And
  Hans-Peter Seidel\\
  Max Planck Institute for Informatics\\
  Saarbrücken, Germany\\
  \texttt{hpseidel@mpi-inf.mpg.de}\\
}

\begin{document}
\maketitle

\begin{figure*}[t]
    \centering
    \input{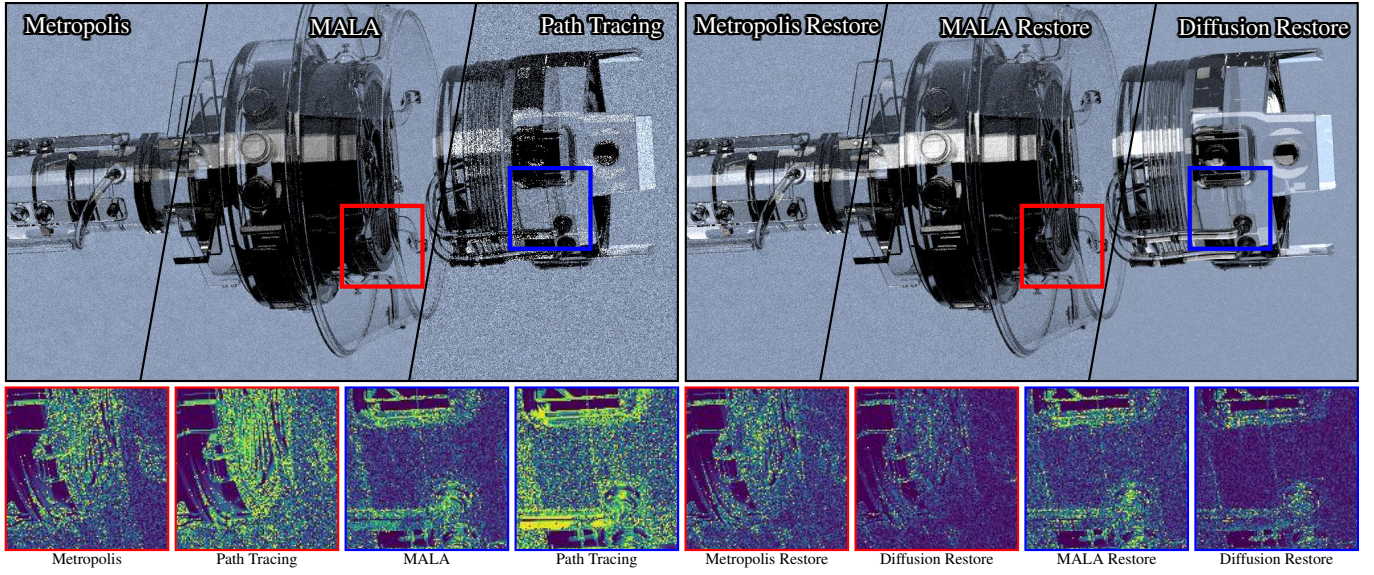}
    
    \captionsetup{skip = 0pt}
    \caption{
        We implement a novel diffusion-based real-time \gls{mcmc} framework on the GPU.
        The rendering shows the \textsc{Transparent Machines} scene
        from \citet{pharr2025scenes} after a rendering time of $1/30$ seconds,
        corresponding to the time budget of an interactive application running
        at 30~\gls{fps}, and compares against current state-of-the-art baselines.
        Further details are provided in \autoref{sec:numerical-study}.
    }
    
    \label{fig:teaser}
\end{figure*}

\begin{abstract}
    We present \emph{Diffusion Restore}, a real-time framework for diffusion-based \gls{mcmc} light transport.
    \gls{mcmc} methods are highly suitable for sampling from complex high-dimensional distributions and estimating integrals over them. In practice, they are often the only viable solution when direct sampling is not possible and alternative methods are either inefficient or cannot be applied due to the structure of the target distribution.
    However, controlling the exploration of the target distribution in \gls{mcmc} methods remains challenging. Efficient exploration requires a balance between \emph{local exploration} and \emph{global discovery}, and local dynamics must rapidly explore individual modes without getting stuck or exhibiting excessive backtracking.
    The problem of \emph{global discovery} has recently been addressed by the introduction of the \emph{Restore} framework. In this work, we build on this framework and focus on improving \emph{local exploration}. We show how to choose diffusion-based local dynamics within the Restore framework while completely avoiding \emph{Metropolis-adjustment}, which often slows down convergence. Furthermore, we model these dynamics as \emph{nonreversible}, introducing momentum in the drift and thereby enabling more directed exploration of the target distribution compared to reversible, random-walk–like dynamics.
    We provide a theoretical justification for the validity of our choice of local dynamics. Empirically, we demonstrate across scenes with diverse light transport characteristics that \emph{Diffusion Restore} outperforms all existing \gls{mcmc} light transport methods and establishes a new state of the art.
    In addition, we present a GPU implementation in ray tracing and compute shaders and achieve real-time frame rates. This demonstrates that Diffusion Restore is not only superior in offline rendering, but also outperforms traditional \acrlong{pt} methods in real-time rendering settings, such as interactive applications and games.
\end{abstract}

% ----------------------------------------------------------------------------------------------------
% introduction
% ----------------------------------------------------------------------------------------------------
\section{Introduction}

\paragraph*{\gls{mcmc} light transport}

Light transport requires the computation of high-dimensional integrals. In practice, these integrals are approximated using \gls{mc} integration. To this end, one considers a scalar nonnegative function (e.g., luminance) derived from the radiometric contribution function induced by the scene. This function can be interpreted as an unnormalized probability density and subsequently used as the \emph{target density} for \gls{mc} sampling.

Traditionally, the image space is stratified into pixels and \gls{mc} integration is performed independently within each pixel. Each pixel estimate is furthermore computed from independently generated samples. While this approach ensures a uniform spread of samples across the image space by construction, it has a major drawback: the computationally expensive generation of samples is performed without regard to their eventual contribution to the estimate. Even if a sample has no significant impact on the estimate, it is still generated without considering the value of the target density at the sampled location.
\acrfull{mcmc} methods offer a way to address this inefficiency. By constructing a \emph{Markov process}, sample generation can be guided to follow the target distribution, thereby concentrating computational effort on regions that contribute most to the estimate.

\paragraph*{Metropolis Light Transport}

\gls{mcmc} methods are highly suitable for sampling from complex high-dimensional distributions. However, controlling the exploration of the target distribution remains challenging. Efficient exploration requires balancing \emph{local exploration} and \emph{global discovery}. On the one hand, a balance between these two aspects must be achieved; on the other hand, the local dynamics must rapidly explore individual modes without getting stuck or repeatedly backtracking.
\citet{veach1997thesis} introduced the \gls{mh} algorithm, a widely used \gls{mcmc} method, to the graphics community. Building on this pioneering work, numerous \gls{mh}-based \gls{mcmc} light transport algorithms have since been proposed. However, all \gls{mh}-based methods share the problem that local explorations tend to random-walk-like behavior and excessive backtracking. This behavior is largely caused by the fact that \gls{mh} dynamics are inherently \emph{reversible} \citep[Definition~1.5.1]{douc2018markov} due to the acceptance/rejection step. While reversibility is beneficial for theoretical analysis, it often leads to slower convergence in practice \citep{neal2004nonreversible}. Furthermore, in light transport applications, global discovery is typically achieved only through an artificial \emph{large step} mechanism that does not adapt to the target density.

\paragraph*{\acrlong{jrlt}}

Recently, \citet{holl2025jrlt} introduced an alternative \emph{continuous-time} \gls{mcmc} framework, known as \emph{Restore}. Restore successfully addresses the problem of balancing global discovery and local exploration by introducing a novel target-density-sensitive \emph{killing} mechanism in place of the artificial large-step mechanism.
Within Restore, local explorations are driven by user-defined \emph{local dynamics}. In the specific instance considered in \citet{holl2025jrlt}, namely \emph{\gls{jrlt}}, these local dynamics are still based on \gls{mh}. As demonstrated in \citet{holl2025jrlt,holl2026rao}, \gls{jrlt} already outperforms all previous \gls{mcmc} light transport methods solely due to its novel treatment of global discovery and establishes the current state of the art.
However, since the local dynamics remain \gls{mh}-based, the speed of local exploration continues to suffer from the limitations induced by reversibility. In this work, we therefore focus on the construction of highly efficient local dynamics that rapidly explore the target distribution while avoiding excessive random-walk-like backtracking.

\paragraph*{Real-time rendering}

In interactive applications such as games, additional constraints arise compared to offline rendering, most notably strict time budgets. The number of samples that can be used to compute a single estimate (i.e., a \emph{frame}) is therefore severely limited. At modern display resolutions and frame rates such as 30 or 60~\gls{fps}, this budget may even fall below one sample per pixel.

Existing approaches address this through techniques such as \emph{adaptive sampling}, \emph{denoising}, and \emph{super-resolution} \citep{bálint2026adaptive}. These methods reconstruct high-quality images from sparse samples or redistribute samples across the image, but they do not fundamentally change the underlying \gls{mc} sampling process itself, which remains based on conventional \gls{pt}.
This raises a natural question: how far can light transport be improved by improving the sampling process itself? Our approach is orthogonal to the aforementioned techniques, including spatiotemporal path reuse methods such as \textsc{ReSTIR} \citep{bitterli2020restir}, as it directly replaces the underlying \gls{mc} sampling method rather than relying solely on downstream reconstruction techniques.

\paragraph*{Our contribution}

We propose \emph{Diffusion Restore}, a diffusion-based framework for both offline and real-time \gls{mcmc} rendering.
Building on the Restore framework introduced in \citet{holl2025jrlt}, we propose to directly employ a \emph{diffusion process} as local dynamics within the Restore algorithm, without any \emph{Metropolis-adjustment}.

More precisely, we employ a \emph{nonreversible Langevin process} (\autoref{sec:langevin-dynamics}), whose dynamics are invariant with respect to the target distribution while introducing a momentum-like rotational component into the drift. Thereby, they enable more persistent and directed exploration of the target distribution while reducing random-walk-like backtracking and can significantly accelerate local mode exploration \citep{eberle2025nonreversible,duncan2017nonreversible,eberle2024relaxation}.

Since Langevin dynamics are generally not exactly simulatable, we effectively employ a continuous-time-embedded discretization as local dynamics. Although this discretization is no longer strictly invariant, exact invariance is not required within Restore, since invariance is enforced through the killing mechanism.
However, the killing rate now contains an additional adjoint term that is computationally expensive to evaluate in practice. We therefore introduce a \emph{modified killing rate} and theoretically analyze the resulting estimator bias (\Cref{sec:killing-rate-modification,sec:bias-analysis}). Furthermore, we demonstrate empirically in \autoref{sec:numerical-study} that the induced bias remains negligible compared to the underlying \gls{mc} error even after prolonged rendering times.

% The results demonstrate that Diffusion Restore establishes a new state of the art in \gls{mcmc}-based offline rendering.

We present a GPU implementation of our algorithm (\autoref{sec:gpu-implementation}) and demonstrate that even under interactive time budgets
%, such as 30~\gls{fps},
the error of Diffusion Restore remains substantially below that of conventional \gls{pt}, which is currently the foundation of practical real-time rendering systems.

% ----------------------------------------------------------------------------------------------------
% related work
% ----------------------------------------------------------------------------------------------------
\section{Related work}\label{sec:related-work}

\paragraph*{Seminal work}

The rendering equation~\citep{kajiya1986rendering} is typically solved using \gls{mc} estimators such as path tracing~\citep{pharr2023pbrt} and its bidirectional variant~\citep{lafortune1996rendering,veach1994bidirectional}. While effective, these approaches sample paths independently of their eventual contribution, which can make certain transport effects difficult to capture efficiently. To address this, \citet{veach1997thesis} introduced \gls{mcmc} into rendering by adapting the \gls{mh} algorithm~\citep{metropolis1953equation,hastings1970monte}. Subsequently, \citet{kelemen2002simple} proposed the \emph{primary sample space} formulation, replacing the highly complex \emph{path space} by a Euclidean parameterization. This significantly simplifies proposal construction and makes the design of efficient sampling strategies more tractable. As a result, the primary sample space formulation has become the standard basis for many \gls{mh}-based rendering methods.

\paragraph*{Diffusion-based \gls{mh}}

A large class of \gls{mh} methods rely on carefully designed proposal kernels. In statistics, such proposals are often derived from discretizations of diffusion processes. Notably, Langevin~\citep{roberts1996mala} and Hamiltonian dynamics~\citep{duane1987hmc} incorporate gradient information to guide proposals and improve local exploration. These ideas have been transferred to rendering: \citet{li2015anisotropic} employed Hamiltonian dynamics with higher-order derivatives, while \citet{luan2020langevin} demonstrated that first-order gradient information already provides an effective trade-off between performance and computational cost.

\paragraph*{\Gls{bdpt}}

\citet{hachisuka2014multiplexed} combined \gls{bdpt} with \gls{mh}, allowing the algorithm to sample across multiple techniques via strategy-dependent proposals. This multiplexed formulation produces estimators analogous to multiple importance sampling and forms the foundation for many modern \gls{mcmc}-based rendering approaches.

\paragraph*{Stratification and global exploration}

Balancing local exploration with global discovery remains a central challenge in \gls{mcmc}-based rendering. \citet{gruson2020stratified} addressed this by distributing multiple chains across predefined strata, thereby improving coverage of the integration domain and stabilizing convergence.

\paragraph*{The Restore framework}

While \gls{mh} has traditionally dominated \gls{mcmc}-based light transport, \citet{wang2021regeneration} introduced \emph{Restore}, a continuous-time \gls{mcmc} framework in which local dynamics are terminated after certain time and restarted from new states. \citet{mckimm2024adaptive} later proposed adaptive regeneration strategies to reduce unnecessary restarts. More recently, \citet{holl2025jrlt} significantly extended the framework by allowing more general local dynamics, state-dependent regeneration mechanisms, and tailoring Restore specifically to light transport simulation.

\paragraph*{Variance reduction}

Beyond stratification and \emph{multiple importance sampling} \citep{veach1994bidirectional}, a range of dedicated variance reduction techniques has been developed for \gls{mh} estimators. Among these, \acrfull{rb} methods \citep{robert2021rao} are particularly relevant. In rendering, a commonly used \gls{rb} technique is \emph{waste-recycling} \citep{ceperley1977waste}, which has been applied in light transport contexts~\citep[Section~4.6]{rioux2020delayed}. A different class of \gls{rb} estimators, referred to as \emph{vanilla \acrlong{rb}}, was introduced by \citet{douc2011vanilla}. While theoretically appealing, these methods incur significant computational overhead in rendering applications. Recently, \citet{holl2026rao} proposed a variant of this technique tailored to light transport that achieves substantial variance reduction while avoiding these costs.

\paragraph*{Real-time path tracing}

Recent work on real-time path tracing focuses on improving efficiency under strict computational budgets. One
% important
direction is \emph{adaptive sampling}, where samples are allocated dynamically based on estimated error \citep{bálint2026adaptive,vogels2018Denoising,kuznetsov2018Deep}. Another
% major
line of work concerns \emph{denoising}, often based on deep learning, which reconstructs high-quality images from low sample counts \citep{
% Mitchell1987Generating,
% lee1990note,
Rushmeier1994Energy,xu2005novel,Overbeck2009Adaptive}. Closely related are \emph{super-resolution} techniques that upscale low-resolution renderings to high-resolution outputs while preserving detail \citep{Kazmierczyk2025joint,Thomas2022Temporally}. These approaches are complementary to the improved \gls{mc} sampling techniques considered in this work.

\paragraph*{Langevin dynamics}

Langevin dynamics have been studied extensively in the literature. Numerous works analyze their general convergence properties \citep{barp2021unifying,chen2024langevin,duncan2017langevin,pagès2023langevin,mou2021langevin,wibisono2018sampling,eberle2023inexact}. The acceleration of convergence via nonreversibility has also received significant attention \citep{bierkens2015nonreversible,eberle2024relaxation,duncan2017nonreversible}. The case most relevant to primary sample space light transport \textemdash\ Langevin dynamics on the torus \textemdash\ was studied by \citet{garcía2017torus}.

% ----------------------------------------------------------------------------------------------------
% Markov chain Monte Carlo
% ----------------------------------------------------------------------------------------------------
\section{Markov chain Monte Carlo}\label{sec:markov-chain-monte-carlo}

Given a finite measure \setword{$\targetdistribution$}{inline:target-distribution}, called \emph{target distribution}, \gls{mcmc} is a recipe for constructing an ergodic \citep{meyn1993markov} Markov process with invariant distribution~$\targetdistribution$.
This Markov process $(\process_\timepoint)_{\timepoint \in \timedomain}$, where \setword{$\timedomain$}{inline:time-domain} is either discrete ($\timedomain=\mathbb N_0$) or continuous ($\timedomain=[0,\infty)$), can subsequently be used to estimate the integral \begin{equation}\label{eq:integral}
    \targetdistribution\secondintegrand:=\int\secondintegrand\dd{\targetdistribution}
\end{equation} of a $\targetdistribution$-integrable function \setword{$\secondintegrand$}{inline:integrand}. In fact, \begin{align}\label{eq:ergodic-theorem}
    \ergodicaverage_\timepoint\secondintegrand:=\frac1\timepoint\left.\begin{cases}\displaystyle\sum_{\prevtimepoint=0}^{\timepoint-1}\secondintegrand(\process_\prevtimepoint)&\text{, if }\timedomain=\mathbb N_0\\\displaystyle\int_0^\timepoint\secondintegrand(\process_\prevtimepoint)\dd{\prevtimepoint}&\text{, if }\timedomain=[0,\infty)\end{cases}\right\}\xrightarrow{\timepoint\to\infty}\targetdistribution\secondintegrand
\end{align}
almost surely for all $\secondintegrand\in\mathcal L^1(\targetdistribution)$ \citep[Theorem~25.6]{kallenberg2021probability}.

\paragraph*{Importance sampling}

In practice, $\targetdistribution$ usually admits a density $\targetdensity$ with respect to a reference measure $\referencemeasure$. That is, \begin{equation}
    \targetdistribution\left(\pointset\right)=\int_\pointset\frac\targetdensity{\targetdensity_\referencemeasure}\dd{\referencemeasure}
\end{equation} for every measurable set $\pointset$, where 
$\targetdensity_\referencemeasure:=\referencemeasure\targetdensity$ is a \emph{normalization constant}.
% \begin{equation}
%     \targetdensity_\referencemeasure:=\int\targetdensity\dd{\referencemeasure}.
% \end{equation}
%
In light transport, $\targetdistribution$ serves as an \emph{importance} distribution and $\targetdensity$ typically proportional to the radiometric contribution (e.g., luminance). Specifically, we are then interested in integrals of the form $\referencemeasure\integrand$ for some integrand $\integrand$ with respect to $\referencemeasure$. By choosing $\secondintegrand=\integrand/\targetdensity$, \autoref{eq:ergodic-theorem} can still be used by noting that $\targetdistribution\secondintegrand=\referencemeasure\integrand$.

% ----------------------------------------------------------------------------------------------------
% Restore framework
% ----------------------------------------------------------------------------------------------------
\section{Restore framework}\label{sec:restore-framework}

In this section, we revisit the Restore framework introduced to the graphics community by \citet{holl2025jrlt} and employ modifications necessary to use the local dynamics described in \autoref{sec:langevin-dynamics}.

Restore is designed to efficiently balance \emph{local exploration} and \emph{global discovery} when sampling from the target distribution $\targetdistribution$. In this work, we consider a special instance of the Restore framework: the \emph{Jump} Restore algorithm introduced in \citet[Section~7]{holl2025jrlt}.

Global discovery in Jump Restore is controlled by a user-defined distribution \setword{$\regenerationdistribution$}{inline:global-dynamics}, which determines the spawn locations of local explorations, while local exploration is governed by a user-defined Markov chain $\markovchain$ induced by a prescribed transition kernel $\transitionkernel$.

To use a discrete-time Markov chain $\markovchain$ as local dynamics in the Restore framework, Jump Restore embeds the dynamics of $\markovchain$ into continuous-time by \emph{holding} each state for a random, almost surely strictly positive duration. More precisely, if the current state is $\point$, then the holding time is assumed to follow an exponential distribution $\operatorname{Exp}(\holdingrate(\point))$, where $\holdingrate$ is a strictly positive \emph{holding rate}. We denote the resulting continuous-time process by $\localprocess$.

Local explorations in Jump Restore are \emph{terminated} (or \emph{killed}) according to a strictly positive \emph{killing rate} $\killingrate$.
Upon termination, the local dynamics are \emph{regenerated} and local exploration is continued at a spawn location drawn from a distribution $\regenerationdistribution$, which is therefore called the \emph{regeneration distribution}.
For our purposes, we require the following slight generalization of \citet[Definition~6.1]{holl2026rao}:

\begin{definition}\label{def:jump-restore-process}
    If $\combinedrate:=\holdingrate+\killingrate$, then the \emph{pure-jump type Markov process} \citep[Chapter~IV]{kallenberg2021probability} with transition kernel \begin{equation}\label{eq:pure-jump-type-transition-rule}
    	\frac{\holdingrate(\point)}{\combinedrate(\point)}\transitionkernel(\point,\;\cdot\;)+\frac{\killingrate(\point)}{\combinedrate(\point)}\regenerationdistribution
    \end{equation} is called the \textbf{Jump Restore process with local dynamics }$\bm\transitionkernel$\textbf{, global dynamics }$\bm\regenerationdistribution$\textbf{, holding rate }$\bm\holdingrate$\textbf{, and killing rate }$\bm\killingrate$.
\end{definition}

By construction, when the current state is $\point$, the Jump Restore process performs a transition according to the local dynamics $\transitionkernel(\point,\;\cdot\;)$ with probability $\holdingrate(\point)/\combinedrate(\point)$. With the complementary probability $\killingrate(\point)/\combinedrate(\point)$, the current local exploration is terminated, and a new spawn location is sampled from the global dynamics $\regenerationdistribution$.

\subsection{Ensuring invariance}

For \gls{mcmc} purposes, the killing rate $\killingrate$ must be chosen such that the Jump Restore process is invariant with respect to a desired target distribution $\targetdistribution$. To state the definition of $\killingrate$ ensuring this invariance,
we need to briefly recap a few functional-analytic concepts.

First of all, recall that the dynamics of a time-homogeneous Markov process can be uniquely characterized by its \emph{generator} \citep[Section~4.1]{ethier2009markov}.
A formal definition of the generator concept is given in \citet[Section~3.2.3]{holl2025jrlt}.
By construction, the generator of the Jump Restore process is given by \citep[Theorem~3.24]{holl2024concatenation} \begin{equation}
    \generator=\localgenerator+\killingrate\globalgenerator
    %\left(\generator\testfunction\right)\left(\point\right)=\left(\localgenerator\testfunction\right)\left(\point\right)+\killingrate\left(\point\right)\left(\globalgenerator\testfunction\right)\left(\point\right),
\end{equation} where, for any bounded measurable function $\testfunction$, \begin{equation}
    \left(\localgenerator\testfunction\right)\left(\point\right)=\holdingrate\left(\point\right)\left(\left(\transitionkernel\testfunction\right)\left(\point\right)-\testfunction\left(\point\right)\right)
\end{equation} is the generator of the continuous-time embedding $\localprocess$ of $\markovchain$, and \begin{equation}
    \globalgenerator\testfunction=\regenerationdistribution\testfunction-\testfunction
\end{equation} is the generator of the global dynamics $\regenerationdistribution$.

Second, for any operator $\operator$ acting on the space of bounded measurable functions, we denote by $\operator^\ast$ the \setword{\emph{adjoint operator}}{inline:adjoint-operator}
% \citep[Section~8.4.1]{renardy2004pde}
\citep[Section~3.2.4]{holl2025jrlt}
with respect to the \emph{duality bracket}
% $\langle\;\cdot\;,\;\cdot\;\rangle$ induced by $\referencemeasure$,
\begin{equation}
    % \left\langle\testfunction_1,\testfunction_2\right\rangle:=\int \testfunction_1\testfunction_2,\dd{\referencemeasure},
    \left\langle\testfunction_1,\testfunction_2\right\rangle:=\referencemeasure\left(\testfunction_1\testfunction_2\right),
\end{equation} defined for all bounded measurable functions $\testfunction_1$ and $\testfunction_2\in L^1(\referencemeasure)$.
A Markov process with generator $\generator$ is $\targetdistribution$-invariant if and only if $\generator^\ast\targetdensity=0$ \citep[Proposition~4.9.2]{ethier2009markov}. In our case,
\begin{equation}
    \generator^\ast\targetdensity=\localgenerator^\ast\targetdensity+\globalgenerator^\ast\left(\killingrate\targetdensity\right),
\end{equation}
and the key is to choose $\killingrate$ such that this expression vanishes.

% We now assume that both the target distribution $\targetdistribution$ and the regeneration distribution $\regenerationdistribution$ admit densities $\targetdensity$ and $\regenerationdensity$, respectively, with respect to a common reference measure $\referencemeasure$. That is, \begin{equation}
%     \begin{split}
%         \targetdistribution\left(\pointset\right)&=\int_\pointset\frac\targetdensity{\targetdensity_\referencemeasure}\dd{\referencemeasure};\\
%         \regenerationdistribution\left(\pointset\right)&=\int_\pointset\frac\regenerationdensity{\regenerationdensity_\referencemeasure}\dd{\referencemeasure}
%     \end{split}
% \end{equation} for every measurable set $\pointset$, where
% \begin{equation}
%     \begin{split}
%         \targetdensity_\referencemeasure&:=\int\targetdensity\dd{\referencemeasure};\\
%         \regenerationdensity_\referencemeasure&:=\int \regenerationdensity\dd{\referencemeasure}.
%     \end{split}
% \end{equation}
We assume that the regeneration distribution $\regenerationdistribution$ admits a density $\regenerationdensity$ with respect to the same reference measure $\referencemeasure$ as $\targetdistribution$, i.e. \begin{equation}
    \regenerationdistribution\left(\pointset\right)=\int_\pointset\frac\regenerationdensity{\regenerationdensity_\referencemeasure}\dd{\referencemeasure}
\end{equation} for every measurable set $\pointset$, where
$\regenerationdensity_\referencemeasure:=\referencemeasure\regenerationdensity$.
% \begin{equation}
%     \regenerationdensity_\referencemeasure:=\int \regenerationdensity\dd{\referencemeasure}.
% \end{equation}

Now, as shown in \citet[Section~6.2]{holl2025jrlt}, the Jump Restore process is $\targetdistribution$-invariant if the killing rate $\killingrate$ is chosen as \begin{equation}\label{eq:invariant-killing-rate}
    \killingrate_\ast=\frac{\localgenerator^\ast\targetdensity}{\targetdensity}+\frac{\regenerationdensity}{\regenerationdensity_\referencemeasure}\frac{\killingrate_0}{\targetdensity},
\end{equation} where $\killingrate_0>0$ must be sufficiently large to ensure $\killingrate_\ast>0$.

\subsection{Approximate killing rate and induced bias}\label{sec:killing-rate-modification}

% If the local dynamics $\transitionkernel$ are already $\targetdistribution$-invariant, then \textemdash\ as noted in the previous section \textemdash\ we have $\localgenerator^\ast\targetdensity=0$,
If the local dynamics $\transitionkernel$ are already $\targetdistribution$-invariant, then necessarily $\localgenerator^\ast\targetdensity=0$,
and thus the first term in the definition \eqref{eq:invariant-killing-rate} of $\killingrate_\ast$ vanishes.

In contrast, if $\transitionkernel$ is not itself $\targetdistribution$-invariant, then $\localgenerator^\ast\targetdensity$ is, in general, an integral operator that is very expensive to evaluate. Since this is incompatible with the interactive computation times we aim for, we omit the first term in \eqref{eq:invariant-killing-rate} and instead choose a \emph{modified} killing rate \begin{equation}\label{eq:modified-killing-rate}
    \killingrate=\frac{\regenerationdensity}{\regenerationdensity_\referencemeasure}\frac{\killingrate_0}{\targetdensity}
\end{equation} for Jump Restore, where $\killingrate_0>0$ can now be chosen arbitrarily. This perturbation of the killing rate \eqref{eq:invariant-killing-rate} clearly induces a bias, rendering the algorithm formally an \emph{inexact} \citep{eberle2023inexact} \gls{mcmc} technique.
In fact, with the modified killing rate \eqref{eq:modified-killing-rate}, we now have
\begin{equation}\label{eq:adjoint-generator-modified-killing-rate}
    \generator^\ast\targetdensity=\localgenerator^\ast\targetdensity+\killingrate_0\globalgenerator^\ast\frac{\regenerationdensity}{\regenerationdensity_\referencemeasure}.
\end{equation}

% That is, we intentionally trade exact invariance for substantially improved exploration efficiency.  However, the
The
key observation is that, in the case relevant to our numerical study in \autoref{sec:numerical-study}, where $\regenerationdensity$ is \emph{constant}, the second term in \autoref{eq:adjoint-generator-modified-killing-rate} vanishes. Consequently, our modification \eqref{eq:modified-killing-rate} of the killing rate introduces a bias that is entirely governed by $\localgenerator^\ast\targetdensity$, i.e., by how far the local dynamics are from being themselves $\targetdistribution$-invariant.

\paragraph*{Impact on estimation}

The bias propagates to estimates of integrals $\targetdistribution\integrand$ as follows. If $\tilde\targetdistribution$ denotes the invariant distribution of the Jump Restore process with the modified killing rate \eqref{eq:modified-killing-rate} and $\testfunction$ is a solution to the \emph{Poisson equation} \citep[Section~4.4]{douc2018markov}, then \begin{equation}
    \tilde\targetdistribution\integrand-\targetdistribution\integrand=\tilde\targetdistribution\left(\left(\killingrate-\killingrate_\ast\right)\globalgenerator\testfunction\right).
\end{equation} Hence, the bias is controlled by \begin{equation}\label{eq:bias}
    \killingrate-\killingrate_\ast=-\frac{\localgenerator^\ast\targetdensity}{\targetdensity},
\end{equation} and therefore depends solely on the choice of local dynamics. We return to this point in \autoref{sec:bias-analysis}.

\subsection{GPU implementation}\label{sec:gpu-implementation}

To enable interactive computation times, we implement Jump Restore on the GPU using ray tracing and compute shaders.

As introduced in \citet{holl2025jrlt}, we refer to the execution of a local exploration up to its termination as a \emph{tour}. We store the state of a tour in the following structure:
\begin{equation}
    \begin{split}
        &\texttt{state}=\{\\
        &\quad\quad\texttt{rng},\\
        &\quad\quad\texttt x,\\
        &\quad\quad\texttt{pixelIndex},\\
        &\quad\quad\texttt p,\\
        &\quad\quad\texttt{f\_over\_p},\\
        &\quad\quad\texttt{killed}\\
        &\},
    \end{split}
\end{equation}
where \texttt{rng} denotes the state of a pseudo-random number generator (we used \emph{Philox4x32-10} \citep{salmon2011parallel} for our implementation), \texttt x is the current state of the tour, and \texttt{pixelIndex} is the associated linearized pixel index (determined by the first two components of \texttt x). Furthermore, \texttt p and \texttt{f\_over\_p} store the values of $\targetdensity$ and $\integrand/\targetdensity$ evaluated at \texttt x, respectively, and \texttt{killed} indicates whether the tour was terminated in the previous call of the shader.

We simulate $\texttt{dispatchCount}$ tours in parallel on the GPU and maintain their states in a global buffer \texttt{stateBuffer} of size \texttt{dispatchCount}. Each entry corresponds to one independent tour that evolves across frames. The \texttt{killed} flag enables efficient regeneration: whenever a tour is 
terminated, it is immediately
regenerated in the next dispatch by sampling a new spawn location from the regeneration distribution $\regenerationdistribution$.
During execution, we keep track of the
total
number of completed tours, denoted by $\texttt{tourCount}$. This quantity is required to form the final estimator of $\referencemeasure\integrand$, as described in \citet[Section~C]{holl2026rao}.

Whenever the scene configuration (e.g., the camera pose or geometry) changes in a way that affects the rendered image, all previously accumulated samples become invalid. In this case, we reinitialize the entire \texttt{stateBuffer} as described in \autoref{alg:diffusion-restore-initialize}. If no such change occurs, the existing tours are continued and updated according to \autoref{alg:diffusion-restore-evolve}. At each step, contributions are accumulated to form the Monte Carlo estimate of the image. In light transport, the target density $\targetdensity$ (e.g., luminance) is usually proportional to the path contribution $\integrand$ and hence does not need to be evaluated separately.

The call to \textsc{LocalDynamics} in \autoref{alg:diffusion-restore-evolve} implements the sampling step $\texttt{state.x}\sim\transitionkernel\left(\texttt{state.x},\;\cdot\;\right)$ and depends on the specific choice of local dynamics, which we describe in \autoref{sec:langevin-dynamics}.

After each dispatch of \autoref{alg:diffusion-restore-initialize} or \autoref{alg:diffusion-restore-evolve}, we perform a separate \emph{resolve} pass (\autoref{alg:diffusion-restore-resolve}). In this pass, the accumulated contributions stored in the buffers are normalized (using $\texttt{tourCount}$) and written to the output texture \texttt{outputTexture}, which represents the current rendered image displayed to the user.

\begin{algorithm}[!ht]
    \caption{\textsc{DiffusionRestoreInitialize}(dispatch index $\texttt{dispatchIndex}\in\mathbb N$, frame index $\timepoint\in\mathbb N_0$)}\label{alg:diffusion-restore-initialize}    
    \begin{algorithmic}[1]
        \State$\texttt{state}=\texttt{stateBuffer}\left[\texttt{dispatchIndex}\right]$;
        \State Seed \texttt{state.generator} using \texttt{dispatchIndex} and $\timepoint$;
        \State Sample $\texttt{state.x}\sim\regenerationdistribution$using \texttt{state.generator};
        \State\Call{Accumulate}{$\texttt{dispatchIndex}, \texttt{state}$};
    \end{algorithmic}

    \begin{algorithmic}[1]
        \Procedure{Accumulate}{$\texttt{dispatchIndex}, \texttt{state}$}:
            % \State$\texttt{state.f}=\integrand\left(\texttt{state.x}\right)$;
            % \State$\texttt{state.p}=\targetdensity\left(\texttt{state.x}\right)$;\AlgCommentLeft{In practice, computed from \texttt{state.f}}
            \State$\texttt{state.f}=\integrand\left(\texttt{state.x}\right)$;
            \State$\texttt{state.pixelIndex}=$ linearized index of the pixel hit by \texttt{state.x[0]};
            \State Sample $\Delta\lifetime_1\sim\operatorname{Exp}\left(\holdingrate\left(\texttt{state.x}\right)\right)$ using \texttt{state.generator};
            \State Sample $\Delta\lifetime_2\sim\operatorname{Exp}\left(\killingrate\left(\texttt{state.x}\right)\right)$ using \texttt{state.generator};
            \If{$\Delta\lifetime_1<\Delta\lifetime_2$:}
                \State$\Delta\lifetime=\Delta\lifetime_1$; 
                \State$\texttt{state.killed}=\texttt{false}$;
            \Else:
                \State$\Delta\lifetime=\Delta\lifetime_2$;
                \State$\texttt{state.killed}=\texttt{true}$;
                \State Atomically increment \texttt{tourCount};
            \EndIf
            \State$\texttt{stateBuffer}\left[\texttt{dispatchIndex}\right]=\texttt{state}$;
            \If{$\texttt{state.p}>0$:}
                \State$\texttt{state.f\_over\_p}=\integrand\left(\texttt{state.x}\right)/\targetdensity\left(\texttt{state.x}\right)$
                \State Atomically add $\Delta\lifetime\cdot\texttt{state.f\_over\_p}$ to $\texttt{accumulationBuffer}\left[\texttt{state.pixelIndex}\right]$;
            \EndIf
        \EndProcedure
    \end{algorithmic}
\end{algorithm}

\begin{algorithm}[t]
    \caption{\textsc{DiffusionRestoreEvolve}($\texttt{dispatchIndex}\in\mathbb N)$}\label{alg:diffusion-restore-evolve}
    \begin{algorithmic}[1]
        \State$\texttt{state}=\texttt{stateBuffer}\left[\texttt{dispatchIndex}\right]$;
        \If{\texttt{!state.killed}:}
            % \State Evolve \texttt{state.x} according to $\transitionkernel\left(\texttt{state.x},\;\cdot\;\right)$
            % \Statex\hspace\algorithmicindent using \texttt{state.generator};
            \State\Call{LocalDynamics}{$\texttt{dispatchIndex}, \texttt{state}$};\AlgCommentLeft{\autoref{alg:diffusion-restore-local-dynamics}}
        \Else:
            \State Re-initialize \texttt{state.x} from $\regenerationdistribution$ using \texttt{state.generator};
        \EndIf
        \State\Call{Accumulate}{$\texttt{dispatchIndex}, \texttt{state}$};
    \end{algorithmic}
\end{algorithm}

\begin{algorithm}[t]
    \caption{\textsc{DiffusionRestoreResolve}($\texttt{pixelIndex}\in\mathbb N$)}\label{alg:diffusion-restore-resolve}
    \begin{algorithmic}[1]
        \State$\texttt{outputTexture}\left[\texttt{pixelIndex}\right]=\displaystyle\frac{\killingrate_0}{\texttt{tourCount}}\cdot\texttt{accumulationBuffer}\left[\texttt{pixelIndex}\right]$;
    \end{algorithmic}
\end{algorithm}

% ----------------------------------------------------------------------------------------------------
% Langevin dynamics
% ----------------------------------------------------------------------------------------------------
\section{Langevin dynamics}\label{sec:langevin-dynamics}

The efficiency with which Restore discovers local modes of the target
distribution $\targetdistribution$ critically depends on the choice of local
dynamics.
In the Jump Restore variants considered by \citet{holl2025jrlt,holl2026rao} the local dynamics still rely on \gls{mh}. While \gls{mh}-based methods ensure exact invariance, they suffer from random-walk-like behavior due to reversibility caused by the accept/reject steps.
In practical applications, including light transport, the underlying state space is often Euclidean. Local exploration in Euclidean spaces can be naturally modeled using \emph{diffusion processes}, which are well suited to describe particle motion through space.
The class of diffusion processes that give rise to time-homogeneous Markov processes is given by \glspl{sde}
% of the form
\begin{equation}\label{eq:sde}
    \dd\localprocess_\timepoint=\drift\left(\localprocess_\timepoint\right)\dd\timepoint+\diffusioncoefficient\left(\localprocess_\timepoint\right)\dd\wienerprocess_\timepoint,
\end{equation} where $\drift$ is referred to as the \emph{drift},
$\diffusioncoefficient$ as the \emph{diffusion coefficient}, and $\wienerprocess$ denotes a \emph{Wiener process} (also known as \emph{Brownian motion}) \citep{karatzas1998brownian,klenke2020probability,pascucci2011pde}.

To ensure efficient local mode discovery, it is desirable that the process $\localprocess$
itself admits $\targetdistribution$ as its invariant distribution. It can be shown that under standard regularity assumptions this is the case if and only if \citep{barp2021unifying} the drift satisfies \begin{equation}\label{eq:langevin-drift}
    \drift=\frac{\left(\covariance+\antisymmetry\right)\nabla\ln\targetdensity+\nabla\cdot\covariance}2
\end{equation} where $\covariance:=\diffusioncoefficient\diffusioncoefficient^\ast$ and $\antisymmetry$ is an antisymmetric matrix. We penalize the target density $\targetdensity$ by a small constant $\epsilon=1\mathrm{e-}8$ to enforce
% strict
positivity. The resulting process is commonly referred to as a \emph{Langevin process}.

\paragraph*{Relation to optimization}

The structural similarity between Langevin dynamics and \emph{stochastic gradient ascent} is not coincidental \citep{ma2015gradient}: the drift term drives the process towards regions of high probability mass, i.e., towards the modes of the target density $\targetdensity$.

\paragraph*{The reversible case}

For $\antisymmetry=0$, the process $\localprocess$ is \emph{reversible}, which yields a number of useful functional-analytic properties and simplifies theoretical analysis. However, in practical applications, reversibility often leads to random-walk–like behavior,
which can slow down convergence due to excessive backtracking \citep{neal2004nonreversible}.

\paragraph*{The nonreversible case}

For $\antisymmetry\neq0$, the drift acquires a rotational component,
effectively introducing \emph{momentum} into the dynamics. This suppresses backtracking behavior and enables more directed exploration of the state space, which in turn accelerates the discovery and traversal of local modes \citep{neal2004nonreversible,chen2013nonreversible,eberle2025nonreversible}.
The advantage of nonreversible Langevin dynamics lies in its more directed and systematic exploration of local modes \textemdash\ without excessive backtracking or random-walk-like behavior. For a Gaussian mode of an example target density, we illustrate the differences between \gls{pt}, Metropolis, \gls{mala}, and Langevin dynamics as used for local exploration in Diffusion Restore.

\begin{figure}[!ht]
    \centering

    \begin{minipage}[t]{.41\textwidth}
        \vspace{0pt}
        \centering

        \begin{subfigure}[t]{.48\linewidth}
            \centering
            \includegraphics[width=\linewidth]{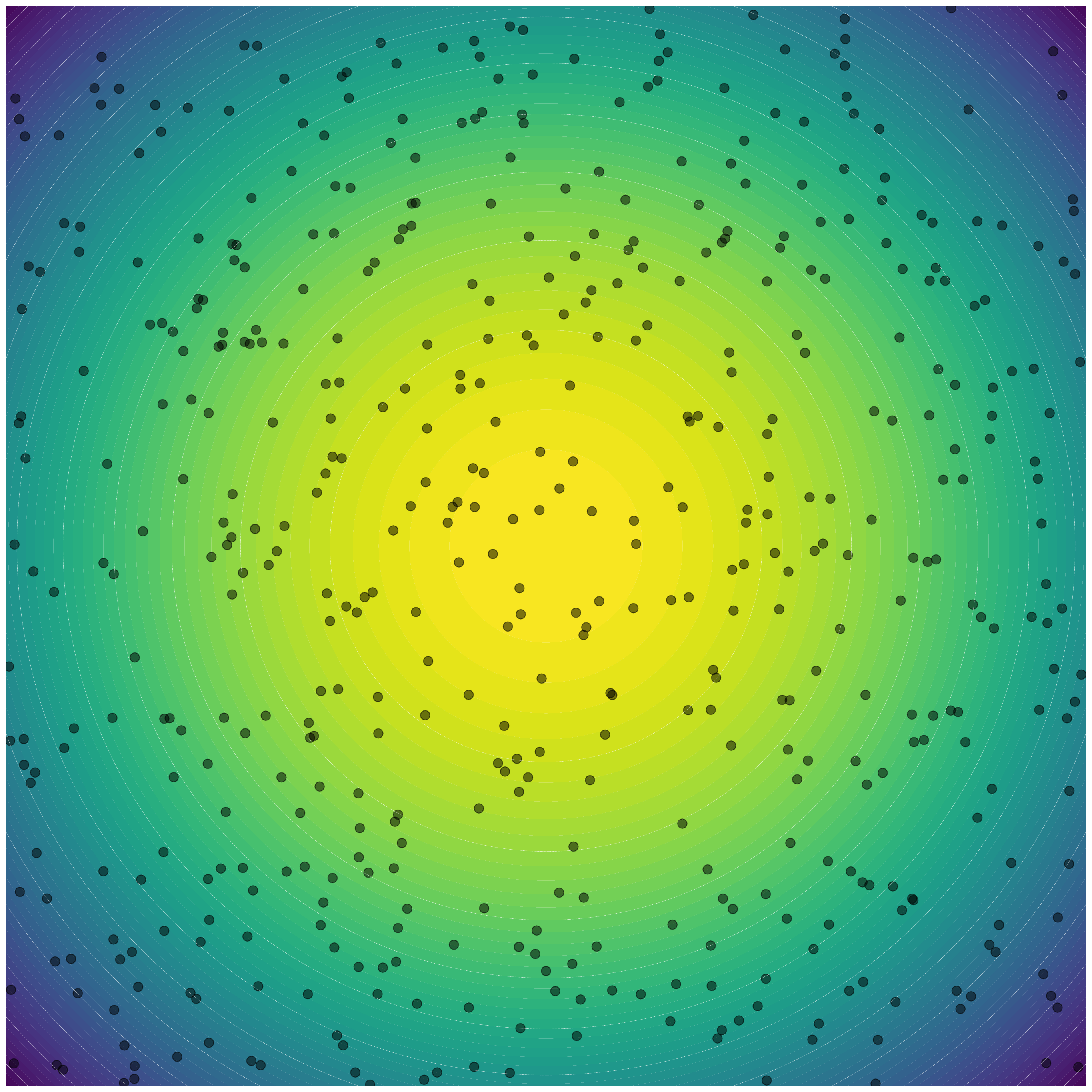}
            \captionsetup{skip = 0pt}
            \caption{Ordinary Monte Carlo}
            \label{fig:exploration-path-tracing}
        \end{subfigure}
        \begin{subfigure}[t]{.48\linewidth}
            \centering
            \includegraphics[width=\linewidth]{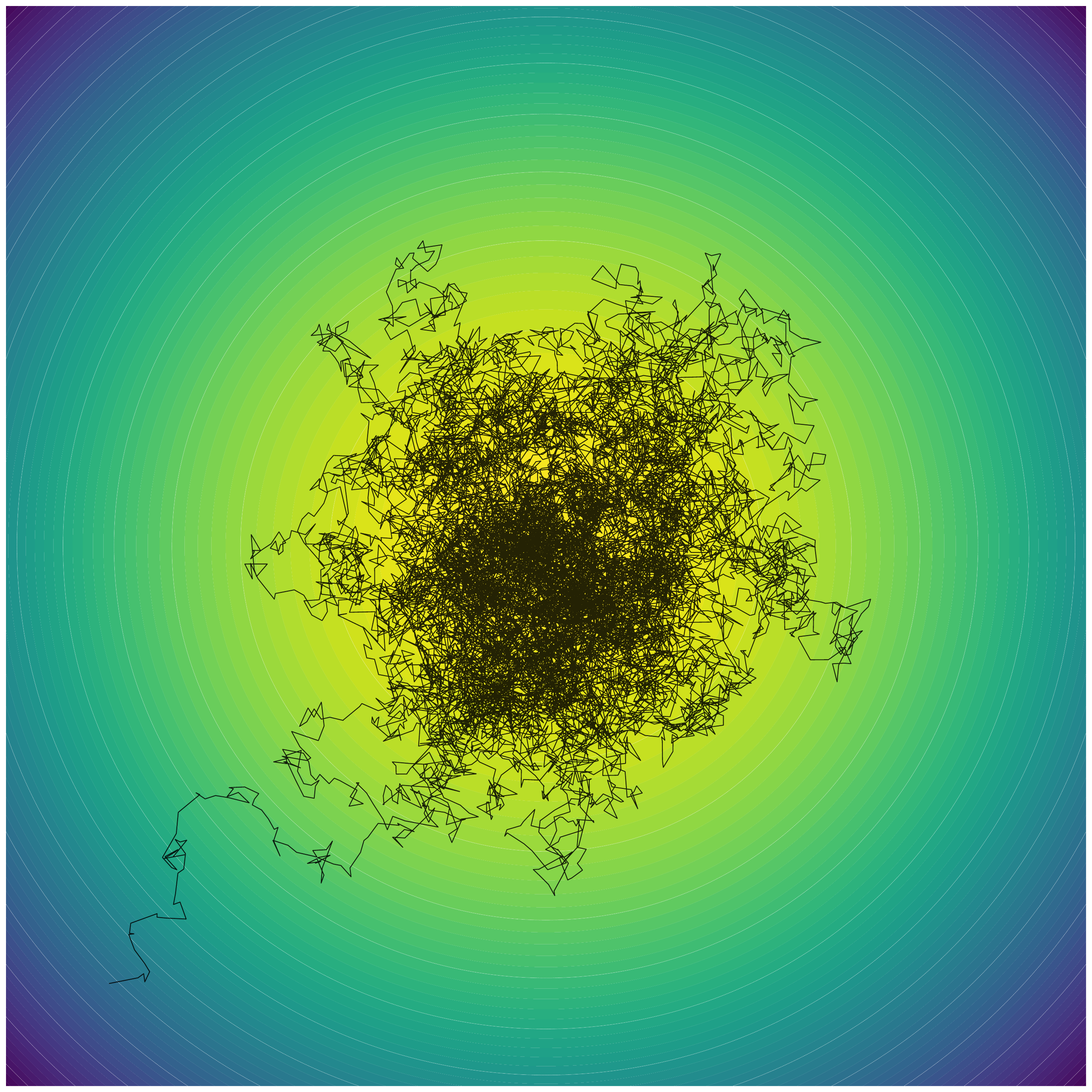}
            \captionsetup{skip = 0pt}
            \caption{\gls{mh}}
            \label{fig:exploration-metropolis}
        \end{subfigure}

        \begin{subfigure}[t]{.48\linewidth}
            \centering
            \includegraphics[width=\linewidth]{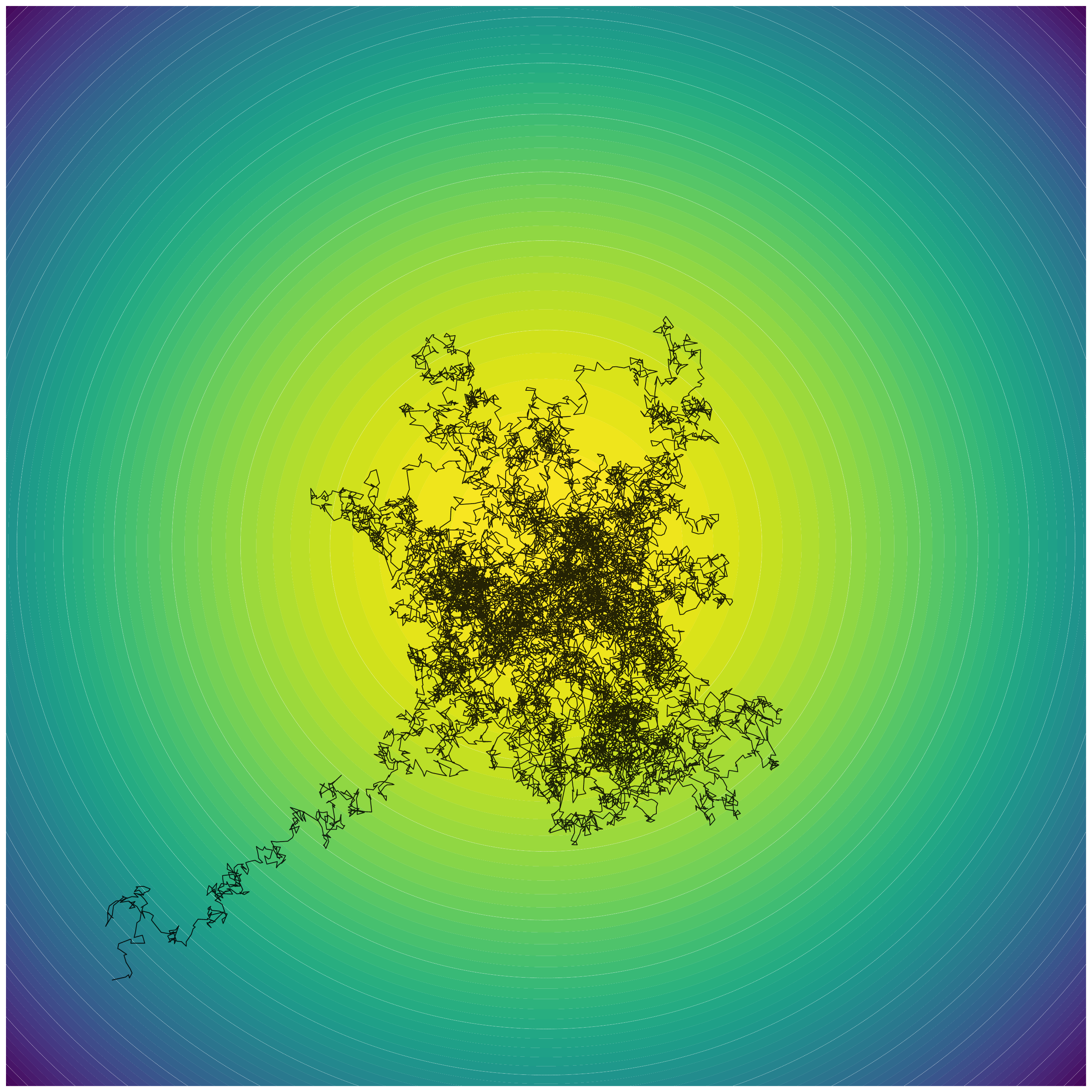}
            \captionsetup{skip = 0pt}
            \caption{\gls{mala}}
            \label{fig:exploration-mala}
        \end{subfigure}
        \begin{subfigure}[t]{.48\linewidth}
            \centering
            \includegraphics[width=\linewidth]{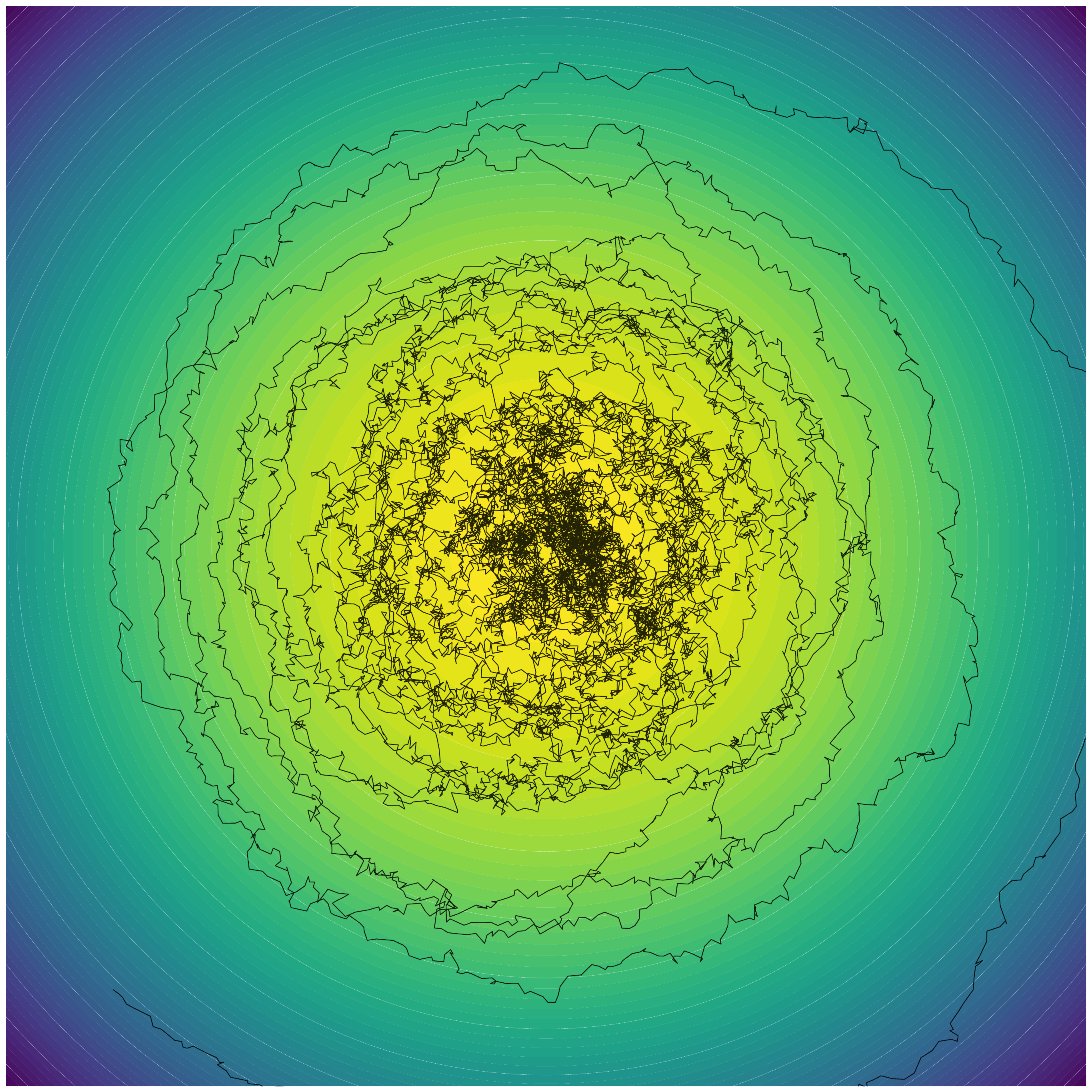}
            \captionsetup{skip = 0pt}
            \caption{Nonreversible Langevin}
            \label{fig:exploration-langevin}
        \end{subfigure}
    \end{minipage}
    \hspace{-.25cm}
    \begin{minipage}[t]{.05\textwidth}
        \vspace{0pt}
        \centering
        \includegraphics[
            height = 7.61cm
        ]{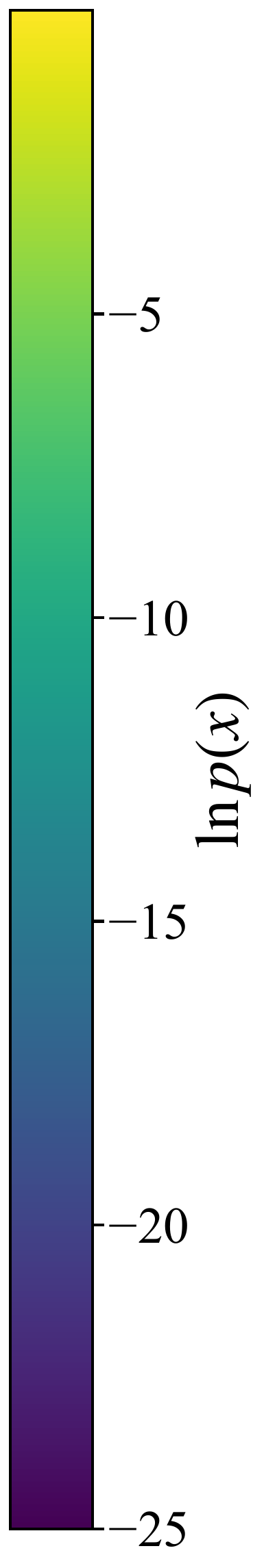}
    \end{minipage}

    \captionsetup{skip = 0pt}
    \caption{
        Exploration of a mode of a target density $\targetdensity$. (a) Ordinary Monte Carlo (i.e., \gls{pt}) generates independent samples. While this ensures uniform coverage and stratification of the domain, many samples fall into low-density regions of the mode. (b) Metropolis sampling adapts the sampling trajectory to the target density and locally to the mode. Due to the reversibility of \gls{mh}-based methods, the resulting exploration exhibits random-walk behavior with frequent backtracking. (c) Metropolis-adjusted Langevin sampling guides proposals toward high-density regions via gradient information. However, the Metropolis-adjustment enforces reversibility, and the resulting trajectory still suffers from the same random-walk and backtracking effects as in (b). (d) Unadjusted Langevin dynamics with an additional nonreversible rotational drift. The rotational component induces a structured exploration that circulates around the mode and consistently moves toward high-density regions, resulting in faster and more structured exploration.
    }
    \label{fig:exploration}
\end{figure}

\subsection{Discrete-time approximation}

Unfortunately, direct simulation of the \gls{sde} in \autoref{eq:sde} is challenging. \emph{Exact sampling} \citep{pollock2016exact,beskos2006exact,roberts2005exact,blanchet2020exact} is only possible under restrictive assumptions on the drift $\drift$ and diffusion coefficient $\diffusioncoefficient$, which are rarely satisfied in practice.

Instead, we must resort to an \emph{inexact} \citep{eberle2023inexact} simulation technique. Similar to the \emph{Euler method} in classical numerical integration, the simplest discretization scheme for an \gls{sde} is the \emph{Euler--Maruyama method} \citep{bencheikh2020eulermaruyama,hutzenthaler2012numeric,kloeden2011numerical}: \begin{equation}\label{eq:euler-maruyama-process}
    \localprocess_{\timepoint+\Delta\timepoint}\approx\localprocess_\timepoint+\Delta\timepoint\drift\left(\localprocess_\timepoint\right)+\sqrt{\Delta\timepoint}\diffusioncoefficient\left(\localprocess_\timepoint\right)\normalvariable,
\end{equation} where $\normalvariable\sim\mathcal N\left(0,I_\dimension\right)$. Apparently, the discretization in \eqref{eq:euler-maruyama-process} induces a Markov chain with transition kernel
\begin{equation}\label{eq:local-dynamics}
    \transitionkernel\left(\point,\;\cdot\;\right):=\mathcal N\left(\point+\Delta\timepoint\drift\left(\point\right),\Delta\timepoint\covariance\left(\point\right)\right).
\end{equation}

\subsection{Bias analysis}\label{sec:bias-analysis}

The invariant measure of the Markov chain with transition kernel \eqref{eq:local-dynamics} coincides with the target distribution $\targetdistribution$ only in the limit $\Delta\timepoint\to0+$ \citep{chen2024langevin,duncan2017langevin,pagès2023langevin,wibisono2018sampling}.
For use within Jump Restore, we choose \begin{equation}\label{eq:diffusion-restore-holding-rate}
    \holdingrate=\frac1{\Delta\timepoint},
\end{equation} as the holding rate so that the continuous-time embedding of $\transitionkernel$ with holding rate $\holdingrate$ is also  $\targetdistribution$-invariant in the limit $\Delta\timepoint\to0+$. More precisely, by the weak error expansion of \citet{talay1990expansion}, \begin{equation}
    \localgenerator^\ast\targetdensity=\Delta\timepoint\correctionoperator^\ast\targetdensity+\mathcal O\left(\Delta\timepoint^2\right),
\end{equation}
where, for any sufficiently regular function $\testfunction$, \begin{equation}
    \correctionoperator\testfunction:=\frac12\operatorname{tr}\left[\left(\drift\otimes\drift\right)\nabla^2\testfunction\right]+\frac12\drift\cdot\nabla\left(\operatorname{tr}\left[\covariance\nabla^2\testfunction\right]\right)+\frac18\operatorname{tr}\left[\covariance\nabla^2\left(\operatorname{tr}\left[\covariance\nabla^2\testfunction\right]\right)\right]
\end{equation} is the first-order correction operator.

\paragraph*{Light transport setting}

In light transport, we operate on the \emph{primary sample space} introduced by \citet{kelemen2002simple}. Formally, this space is a torus $\mathbb T^\dimension=\mathbb R^\dimension/\mathbb Z^\dimension$ for some $\dimension\in\mathbb N$, and is therefore compact. Consequently, both $\correctionoperator^\ast\targetdensity$ and $\targetdensity$ are bounded and hence \begin{equation}\label{eq:adjoint-bias}
    \frac{\localgenerator^\ast\targetdensity}{\targetdensity}=\mathcal O\left(\Delta\timepoint\right).
\end{equation} Thus, according to \autoref{eq:bias}, using the local dynamics $\transitionkernel$ in \autoref{eq:local-dynamics} together with the modified killing rate \eqref{eq:modified-killing-rate} in the Jump Restore algorithm yields a theoretical estimate bias of order $\mathcal O(\Delta\timepoint)$.

\paragraph*{Vanishing practical bias}

% In our numerical study (\autoref{sec:numerical-study}), $\Delta\timepoint$ is chosen sufficiently small to ensure that the bias is dominated by Monte Carlo variance for all practical rendering budgets.
In our numerical study (\autoref{sec:numerical-study}), $\Delta\timepoint$ is chosen sufficiently small such that the resulting bias is empirically negligible compared to \gls{mc} variance.
In \Cref{fig:bias}, we illustrate using a representative target density typical for light transport that \eqref{eq:adjoint-bias} becomes negligible \textemdash\ and often effectively vanishes \textemdash\ for sufficiently small values of $\Delta\timepoint$.

\subsection{The Diffusion Restore process}

Summarizing our construction, we arrive at the following definition:

\begin{definition}
    We call the Jump Restore process with local dynamics \eqref{eq:local-dynamics}, global dynamics $\regenerationdistribution$, holding rate \eqref{eq:diffusion-restore-holding-rate} and killing rate \eqref{eq:modified-killing-rate} the \textbf{Diffusion Restore process with target distribution }$\bm\targetdistribution$\textbf{ and regeneration distribution }$\bm\regenerationdistribution$.
\end{definition}

Here, $\regenerationdistribution$ may in principle still be \emph{any} distribution on $\mathbb T^\dimension$. However, for our numerical study in \autoref{sec:numerical-study}, we restrict ourselves to the uniform distribution on $\mathbb T^\dimension$. Moreover, we choose a constant scalar diffusion coefficient $\diffusioncoefficient\in(0,\infty)$ (yielding $\covariance=\diffusioncoefficient^2I_\dimension$) and define \begin{equation}
    % \antisymmetry=\rotationscale\underbrace{\operatorname{diag}\left(\antisymmetry_{\textnormal{block}},\ldots,\antisymmetry_{\textnormal{block}}\right)}_{=:\;\tilde\antisymmetry}
    \antisymmetry=\rotationscale\underbrace{\operatorname{diag}\left(\antisymmetry_{\textnormal{block}},\ldots,\antisymmetry_{\textnormal{block}}\right)}_{=:\;\tilde\antisymmetry}
\end{equation} for some $\rotationscale\in(0,\infty)$
, where \begin{equation}
    \antisymmetry_{\textnormal{block}}:=\begin{pmatrix}0&1\\-1&0\end{pmatrix}.
\end{equation}
.
This construction introduces a rotational component to the dynamics. The two-dimensional block structure reflects the fact that most path sampling subroutines consume pairs of random numbers \citep{pharr2023pbrt}. The parameter $\rotationscale$ is a scaling factor controlling the intensity of the rotational drift.

Reflecting the block structure, we identify $\mathbb T^\dimension\simeq[0,1)^\dimension\simeq([0,1)^2)^\reduceddimension$, where we assume without loss of generality that $\dimension\in2\mathbb N$ and set $\reduceddimension:=\dimension/2$. For implementation, it is convenient to reparameterize the dynamics in terms of quantities that directly control the magnitude of the updates. We combine the step size $\Delta\timepoint$ and the diffusion coefficient $\diffusioncoefficient$ into the standard deviation of the Gaussian increments, 
% \begin{equation}
%     \texttt{stddev}:=\sqrt{\Delta\timepoint}\diffusioncoefficient,
% \end{equation}
$\texttt{stddev}:=\sqrt{\Delta\timepoint}\diffusioncoefficient$,
and similarly absorb the step size into the rotational component via
% \begin{equation}
%     % \tilde\rotationscale:=\frac{\Delta\timepoint\rotationscale}{2}.
%     \tilde\rotationscale:=\left(\Delta\timepoint\rotationscale\right)/2.
% \end{equation}
$\tilde\rotationscale:=\left(\Delta\timepoint\rotationscale\right)/2$.
With this parameterization, the drift term in \autoref{eq:langevin-drift} becomes
\begin{equation}
    \Delta\timepoint\frac{\left(\covariance+\antisymmetry\right)\nabla\ln\targetdensity}2=\left(\frac{\texttt{stddev}^2}2+\tilde\rotationscale\tilde\antisymmetry\right)\nabla\ln\targetdensity.
\end{equation}
The complete evolution step of these local dynamics is summarized in \autoref{alg:diffusion-restore-local-dynamics}.

\begin{algorithm}[t]
    \caption{\textsc{LocalDynamics}(\texttt{state})}\label{alg:diffusion-restore-local-dynamics}
    \begin{algorithmic}[1]
        \State $\texttt{score}=\nabla\ln\targetdensity\left(\texttt{state.x}\right)$;\AlgCommentLeft{$\in(\mathbb R^2)^\reduceddimension$}
        \For{$\left(\dimensionindex=0;\dimensionindex<\reduceddimension;\textnormal{++}\dimensionindex\right)$:}
            \State Sample $\xi\sim\mathcal N(0,I_2)$;\AlgCommentLeft{$I_2$ denotes identity matrix of size $2$}
            \State$\texttt{state.x}\left[\dimensionindex\right]\mathrel+=\frac{\texttt{stddev}^2}2\ast\texttt{score}\left[\dimensionindex\right]+\Call{rotate}{\texttt{score}\left[i\right]}+\texttt{stddev}\cdot\xi;$
            \State$\texttt{state.x}\left[\dimensionindex\right]\mathrel-=\lfloor\texttt{state.x}\left[\dimensionindex\right]\rfloor$;\AlgCommentLeft{Toroidally wrap back to $\mathbb T^2$}
        \EndFor
    \end{algorithmic}

    \begin{algorithmic}[1]
        \Function{rotate}{$v\in\mathbb R^2$}
            \State\Return$\tilde\rotationscale\left(v_2,-v_1\right)$;
        \EndFunction
    \end{algorithmic}
\end{algorithm}

% ----------------------------------------------------------------------------------------------------
% numerical study
% ----------------------------------------------------------------------------------------------------
\section{Numerical study}\label{sec:numerical-study}

\paragraph*{Setup}

We use \textsc{Slang} \citep{slang}, together with its cross-platform graphics layer \textsc{GFX} \citep{gfx} to support multiple backends, to implement Diffusion Restore. \textsc{Slang} enables the computation of gradients via \emph{automatic differentiation} \citep{walther2008evaluating}.

Within the same codebase, we additionally implemented ordinary \gls{pt} as a reference method. Furthermore, we implemented the \gls{mcmc}-based methods studied in \citet{holl2025jrlt,holl2026rao}, namely \emph{Metropolis}, \emph{\gls{mala}}, \emph{Metropolis Restore}, and \emph{\gls{mala} Restore}. While \gls{pt}-based sampling is widely regarded as the current standard for real-time and interactive applications, Metropolis Restore and \gls{mala} Restore represent the state of the art among \gls{mcmc}-based offline methods. Hence, our selection covers all relevant baselines.
% The purely \gls{mh}-based methods Metropolis and \gls{mala} are included for comparison, as they form the foundation of traditional \gls{mcmc} approaches but have not been widely adopted in practice due to inherent limitations of \gls{mh}.

Importantly, all methods share the same underlying path sampling routines; they differ exclusively in the high-level sampling strategy (i.e., independent \gls{mc} versus \gls{mcmc} sampling), rather than in the construction of light transport paths.

The GPU implementations of Metropolis Restore and \gls{mala} Restore largely follow \Cref{alg:diffusion-restore-initialize,alg:diffusion-restore-evolve,alg:diffusion-restore-resolve}. The only differences are that the holding rate $\holdingrate$ is fixed to $1$ (as described in \citet[Section~7.1]{holl2025jrlt}) and that the local dynamics correspond to Metropolis or \gls{mala} transitions (as described in \citet[Section~8]{holl2025jrlt}).

The \gls{mh}-based methods Metropolis and \gls{mala} require, upon scene changes, an additional \emph{bootstrap} pass besides the initialize pass to compute the normalization constant $\targetdensity_\referencemeasure$ of the target density. This bootstrap pass consists of a single dispatch call of image resolution size and uses ordinary \gls{pt}.
For the \gls{pt} implementation, as well as all \emph{initialize} and \emph{evolve} passes of Metropolis, \gls{mala}, Metropolis Restore, \gls{mala} Restore, and Diffusion Restore, we issue one dispatch call of image resolution size per frame, ensuring that all methods generate the same number of samples per frame.

\paragraph*{Parameter choices}

Parameters were chosen based on an ablation study balancing exploration speed and discretization bias.

For Metropolis and Metropolis Restore, we used a \texttt{stddev} of $1\mathrm{e-}2$, while for the Langevin-based methods \gls{mala}, \gls{mala} Restore, and Diffusion Restore, we used $\texttt{stddev}=5\mathrm{e-}3$. For the \gls{mh}-based methods Metropolis and \gls{mala}, we additionally employed a \emph{large step probability} (see \citet[Section~5]{holl2025jrlt}) of $0.3$.% These parameter choices were found to be near-optimal in an ablation study.

For all Restore variants, we set the killing rate parameter $\killingrate_0$ such that each tour performs, on average, $m$ Euler--Maruyama steps before termination. This is achieved by choosing 
\begin{equation}\label{eq:killing-rate-constant}
    % \killingrate_0=\frac{\targetdensity_\referencemeasure}{m\Delta\timepoint}.
    \killingrate_0=\targetdensity_\referencemeasure/\left(m\Delta\timepoint\right).
\end{equation}
% $\killingrate_0=\targetdensity_\referencemeasure/\left(m\Delta\timepoint\right)$.
In all reported experiments, we use $m=64$. In practive, if $\targetdensity_\referencemeasure$ is unavailable, the initialize pass can use an arbitrary $\killingrate_0>0$, and $\targetdensity_\referencemeasure$ can be estimated from this pass. The estimate can then be used to set $\killingrate_0$ according to \eqref{eq:killing-rate-constant} for all subsequent evolve passes. Note, however, that this particular choice of $\killingrate_0$ is not required; it is introduced here only to provide better control over the expected length of local tours.
For Diffusion Restore, we set $\Delta\timepoint=1\mathrm{e-}5$. The theoretically predicted bias of $\mathcal O(\Delta\timepoint)$ is negligible in practice, as confirmed empirically in our results (see also the \emph{bias assessment} paragraph below and our preceding discussion on the vanishing bias in \autoref{sec:bias-analysis}).

% We implemented all \gls{mcmc}-based methods in two variants: one operating on the full primary sample space, and one separating direct and indirect illumination.
% %
% The latter variant is motivated by prior work, which often applies \gls{mh}-based methods primarily to indirect illumination while handling direct illumination via conventional sampling techniques, as direct illumination can typically be sampled efficiently without \gls{mcmc} \citep{lehtinen2013gradient,bashford2021ensemble}. 
% %
% In our experiments, however, we observed that applying \gls{mcmc} methods to both direct and indirect illumination consistently yielded superior metrics. We therefore report results exclusively for the full-dimensional variants.
In contrast to prior work~\citep{lehtinen2013gradient,bashford2021ensemble}, we employ \gls{mcmc} sampling for \emph{both} direct and indirect illumination, as our experiments consistently yielded improved metrics when applying \gls{mcmc} on the entire state space.

% Further details can be found in our code, which will be made available to the community upon acceptance.

\paragraph{Test scenes}

We evaluated a diverse set of scenes exhibiting different light transport
characteristics. For the purpose of the numerical evaluation, all experiments
were conducted at a rendering resolution of $1024\times768$ to enable
extensive equal-rendering-time and equal-sample-count comparisons across
methods and scenes. The scenes were drawn from multiple sources:
\textsc{Country Kitchen},
% \textsc{Water Caustic},
\textsc{Glass of Water}, \textsc{Salle de bain}, and
\textsc{Veach, Ajar} from \citet{bitterli2016resources};
\textsc{Torus} from \citet{lmc}; \textsc{Swimming Pool}
from \citet{rioux2020delayed}; and \textsc{Transparent Machines}
from \citet{pharr2025scenes}.
We converted all scene descriptions into our rendering system. For this evaluation, we support the \textsc{pbrt-v4} \citep{pharr2023pbrt} materials \textsc{conductor}, \textsc{dielectric}, and \textsc{diffuse}. As light sources, we exclusively use \textsc{pbrt-v4}'s \texttt{DiffuseAreaLight}.

% For the purpose of this submission, texture sampling was disabled. This, however, is not a limitation of Diffusion Restore. \textsc{Slang} provides example applications \citep{nvidia2023differentiable} demonstrating how texture-based computations can be made differentiable.

\paragraph*{Limitations}

For the purpose of this submission, texture sampling was disabled. Existing differentiable rendering systems such as \textsc{Mitsuba Renderer 3} \citep{jakob2022mitsuba3} focus on derivatives with respect to scene parameters rather than primary sample space coordinates. Supporting the latter, while simultaneously targeting a GPU implementation, required a custom differentiable renderer, which at the current stage does not yet support all features commonly found in production renderers, such as texture sampling. This is, however, not a limitation of Diffusion Restore itself. \textsc{Slang} provides example applications \citep{nvidia2023differentiable} demonstrating differentiable texture-based computations.

\paragraph*{Error metrics}

% We evaluate several quantitative metrics: the $L^1$-error (i.e., \gls{mae}), $L^2$-error (i.e., \gls{mse}), \gls{mrse}, and \gls{mape}. Reference images are generated using \gls{pt} with $2^{24}$ \gls{spp}.
We evaluate the $L^1$-error (\acrshort{mae}), $L^2$-error (\acrshort{mse}), \acrshort{mrse}, and \acrshort{mape}. Reference images use \gls{pt} with $2^{24}$ \gls{spp}.
To assess potential bias, all metrics are averaged over 100 independent realizations. For each realization, a different random seed is used, and the same set of 100 seeds is shared across all methods. This ensures that any observed differences are solely due to the respective sampling and exploration strategies.
All methods are run for up to at least 360 rendering seconds and $2^{18}$ \gls{spp}. All measurements reached an \gls{mse} below $1\mathrm{e-}5$.

\begin{figure}
    \centering
    \includegraphics[width=\linewidth]{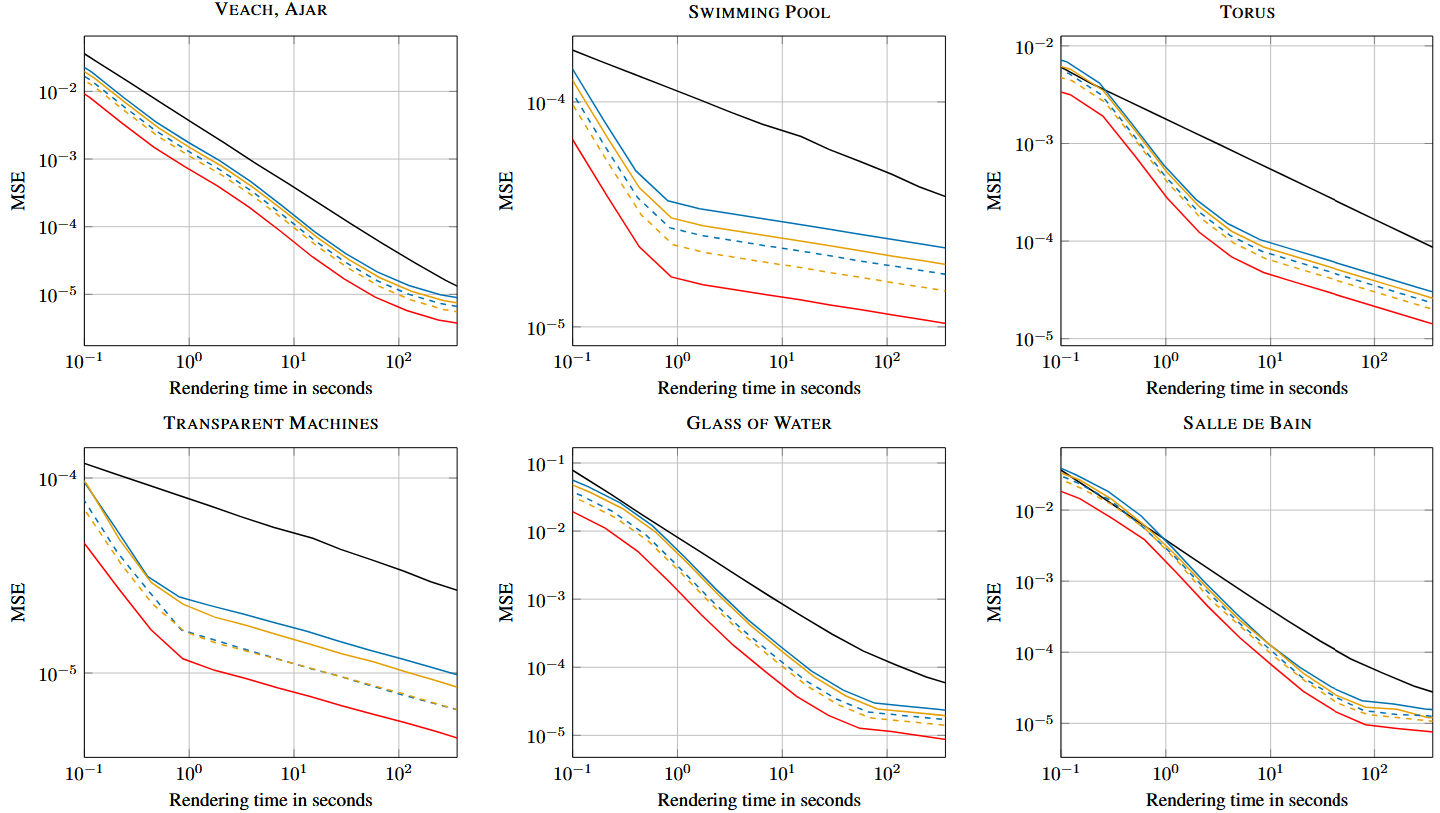}
    \caption{
        \gls{mse} as a function of rendering time
        (up to 360 seconds, averaged over 100 realizations)
        for the scenes shown in
        \Cref{fig:teaser,fig:pool,fig:torus,fig:bathroom,fig:glass-of-water,fig:veach-ajar}.
        \gls{pt} exhibits a linear decay in log-scale due to independent sampling,
        while \gls{mcmc}-based methods can show curvature because correlation effects
        lead to a non-constant convergence rate.
        All \gls{mcmc} methods converge faster than \gls{pt};
        gradient-based variants (\gls{mala}, \gls{mala} Restore)
        further improve convergence due to guided sampling,
        and Diffusion Restore achieves the fastest convergence
        due to momentum and the absence of a Metropolis-adjustment.
    }
    \label{fig:error}
\end{figure}

Plots of the \gls{mse} over rendering time for the scenes shown in \Cref{fig:teaser,fig:pool,fig:torus,fig:bathroom,fig:glass-of-water,fig:veach-ajar} are provided in \autoref{fig:error}. Additional plots showing all metrics as functions of rendering time (\Cref{sec:equal-rendering-time}) and \gls{spp} (\Cref{sec:equal-sample-count}), as well as error tables at fixed rendering times and fixed \gls{spp}, can be found in the appendix.

As the results show, all \gls{mcmc} sampling methods exhibit accelerated convergence compared to \gls{pt}. The gradient-based methods, \gls{mala} and \gls{mala} Restore, achieve improved convergence despite their higher computational complexity, due to more effective sample guidance toward high-probability regions. Finally, our Diffusion Restore sampler further accelerates convergence, owing to the introduced momentum and the absence of a Metropolis-adjustment. Due to the nature of global discovery in the \gls{mh} methods Metropolis and \gls{mala}, the renderings still exhibit a significant amount of bias at early time budgets such as $1/30$ seconds of computation. While bias-reduction techniques could in principle mitigate this effect, they would consume part of the same fixed computational budget and thus do not improve performance under equal-time constraints. These observations are consistent across all evaluated scenes.

\paragraph*{Bias assessment}

In all reported experiments, we do not observe any bias-dominated regime. Even at long rendering times (up to 360 seconds and $2^{18}$ \gls{spp}), the error metrics continue to decrease without saturation and rapidly approach $0$. Especially in real-time applications, where sample budgets are significantly lower, the discretization bias is negligible compared to \gls{mc} error.

\paragraph*{Visual assessment}

A comparison across all methods is shown in \Cref{fig:teaser}. In \Cref{fig:pool,fig:torus,fig:bathroom,fig:glass-of-water,fig:veach-ajar}, we focus on Diffusion Restore, \gls{pt}, and \gls{mala} Restore \textemdash\ the current state of the art for offline \gls{mcmc} light transport \citep{holl2026rao}. Comprehensive comparisons across all methods and scenes will be made available through an interactive viewer accompanying this paper, which is scheduled for public release later in 2026. All images are rendered with an equal computation time of $1/30$ seconds, corresponding to the per-frame budget of a 30~\gls{fps} application.
In addition, the interactive viewer will provide videos comparing \gls{pt}, Metropolis, and Diffusion Restore side by side, where Metropolis serves as a classical baseline, having long been the standard \gls{mcmc} rendering method. For these videos, all methods are rendered along the same pre-recorded camera trajectory, parameterized by arc length to ensure constant motion, and evaluated at a fixed budget of 30~\gls{fps}. This setup enables a fair comparison of visual quality under identical computational constraints.

We note that our optimized GPU implementation of \gls{mala} Restore already provides highly competitive performance. As a result, gains achieved by Diffusion Restore may appear marginal in certain scenes, particularly at short rendering times. Nevertheless, these improvements accumulate over time and become increasingly significant as the dimensionality of the sampling problem grows, where improved local exploration lead to progressively larger advantages.

\paragraph*{Hardware resources}

All experiments were conducted on a desktop system equipped with an AMD Ryzen 9 7950X3D CPU, an NVIDIA GeForce RTX 4080 Super GPU (16\,GB GDDR6X, Gainward Python III OC), and 96\,GB DDR5-6400 CL32 memory (Kingston Fury Renegade, 2$\times$48\,GB). The GPU was used for all rendering and \gls{mcmc} computations, while the CPU handled host-side orchestration.

% ----------------------------------------------------------------------------------------------------
% conclusion
% ----------------------------------------------------------------------------------------------------
\section{Conclusion}\label{sec:conclusion}

% While MH ensures exact invariance, it suffers from random-walk behavior due to reversibility and rejection. In contrast, Diffusion Restore sacrifices exactness at the level of the local dynamics but achieves significantly improved exploration, leading to faster variance reduction.

% In this work, we
We
introduced Diffusion Restore, an \gls{mcmc} light transport method that consistently outperforms prior \gls{mcmc}-based state-of-the-art offline rendering techniques across all considered metrics.
Furthermore, we presented a GPU implementation and provided an empirical evaluation demonstrating that Diffusion Restore also achieves superior performance under real-time constraints compared to all competing methods \textemdash\ including conventional, \gls{pt}, which is the standard sampling approach in interactive applications such as games.

% ----------------------------------------------------------------------------------------------------
% future work
% ----------------------------------------------------------------------------------------------------
\section{Future work}\label{sec:future-work}

Exploring additional nonreversible drift components $\antisymmetry$ has the potential to further accelerate convergence.

Beyond improving local exploration, another key aspect of the Restore framework that remains largely unexplored is the design of the global dynamics. Instead of sampling regeneration points uniformly, it may be beneficial \textemdash\ especially in interactive settings \textemdash\ to exploit information about previously discovered modes when constructing the regeneration distribution $\regenerationdistribution$ for subsequent frames. In this context, generative approaches may provide a promising direction.

% Beyond improving local exploration, another key aspect of the Restore framework that remains largely unexplored is the design of the global dynamics. In the current formulation, accumulated samples become invalid as soon as the scene configuration changes (e.g., due to camera motion), requiring a full reinitialization of the process. Instead of sampling regeneration points uniformly, it may therefore be beneficial \textemdash\ especially in interactive settings \textemdash\ to leverage information from previously rendered frames when constructing the regeneration distribution $\regenerationdistribution$ for subsequent frames. In this context, generative approaches could be used to learn a suitable regeneration distribution conditioned on past observations, thereby enabling a more informed initialization of local explorations.

% ----------------------------------------------------------------------------------------------------
% bibliography
% ----------------------------------------------------------------------------------------------------
\bibliographystyle{ACM-Reference-Format}
\bibliography{bibliography}

\newpage

% ----------------------------------------------------------------------------------------------------
% results
% ----------------------------------------------------------------------------------------------------

\clearpage

\begin{figure*}
    %!TEX root = ../main.tex

% Rotated two-line method label (for right margin of table)
% #1 first line, #2 second line
\newcommand{\RotTableMethodLabel}[2]{%
    \smash{%
        \raisebox{0.5\height}{% optisch zentrieren; bei Bedarf feinjustieren
            \rotatebox[origin = c]{90}{%
                \tiny\begin{tabular}{@{}l@{}}
                    \hspace{.6cm}#1\\[.1ex]
                    \hspace{.6cm}#2
                \end{tabular}%
            }%
        }%
    }%
}

% #1 = x position
% #2 = y position
% #3 = anchor
% #3 = first line
% #4 = second line
\newcommand{\MethodLabelAt}[5]{%
  \node[anchor = #3] at (#1, #2) {%
    \shortstack[l]{%
      \contour{black}{\small\textcolor{white}{#4}}\\[-0.2ex]%
      \contour{black}{\small\textcolor{white}{#5}}%
    }%
  };%
}

% Two-line tiny label for tables (left-aligned)
\newcommand{\TableMethodLabel}[2]{%
  \begin{tabular}{@{}l@{}}%
    \tiny #1\\[-0.1ex]%
    \tiny #2%
  \end{tabular}%
}

% \pgfmathsetmacro\width{1024}
% \pgfmathsetmacro\height{768}
\pgfmathsetmacro\width{1024}
\pgfmathsetmacro\height{576}
\pgfmathsetmacro\aspectratio{16 / 9}

\newcommand\scene{veach_ajar}

\newcommand\firstmethod{Path_Tracing}
\newcommand\secondmethod{MALA_Restore}
\newcommand\thirdmethod{Diffusion_Restore}
\newcommand\firstmethoddisplaynamea{Path Tracing}
\newcommand\firstmethoddisplaynameb{\vphantom|}
\newcommand\secondmethoddisplaynamea{MALA Restore}
\newcommand\secondmethoddisplaynameb{\vphantom|}
\newcommand\thirdmethoddisplaynamea{Diffusion Restore}
\newcommand\thirdmethoddisplaynameb{\vphantom|}

\pgfmathsetmacro\croplength{.14}
\pgfmathsetmacro\firstcropxmin{.33}
\pgfmathsetmacro\firstcropymin{.21}
\pgfmathsetmacro\secondcropxmin{.51}
\pgfmathsetmacro\secondcropymin{.21}
\newcommand\firstcropcolor{red}
\newcommand\secondcropcolor{blue}

\newcommand{\SplitThreeImages}[3]{
    \begin{scope}
       \clip (0, 0) -- (.2 * \aspectratio, 0) -- (.3 * \aspectratio, 1) -- (0, 1) -- cycle;
        \path[fill overzoom image = results/#1.jpg] (0, 0) rectangle (\aspectratio, 1);
    \end{scope}
    \begin{scope}
       \clip (.2 * \aspectratio, 0) -- (0.6 * \aspectratio, 0) -- (0.7 * \aspectratio, 1) -- (0.3 * \aspectratio, 1) -- cycle;
        \path[fill overzoom image = results/#2.jpg] (0, 0) rectangle (\aspectratio, 1);
    \end{scope}
   \begin{scope}
       \clip (.6 * \aspectratio, 0) -- (\aspectratio, 0) -- (\aspectratio, 1) -- (0.7 * \aspectratio, 1) -- cycle;
       \path[fill overzoom image = results/#3.jpg] (0,0) rectangle (\aspectratio, 1);
   \end{scope}
   \begin{scope}
        \draw[black, thick] (0, 0) rectangle (\aspectratio, 1);
        \draw[draw=black,thick] (.2 * \aspectratio, 0) -- (.3 * \aspectratio, 1);
        \draw[draw=black,thick] (.6 * \aspectratio, 0) -- (.7 * \aspectratio, 1);
   \end{scope}
}

%   - #1: scene name (without extension; image is results/#1.jpg)
%   - #2: crop x (normalized, 0–1; 0=left)
%   - #3: crop y (normalized, 0–1; 0=bottom; note: image origin is lower left)
%   - #4: crop length (normalized)
%   - #5: color (for drawing the outline)
\newcommand{\CropImage}[5]{%
    \begingroup
        % Convert normalized crop to pixel values
        \pgfmathsetmacro\length{#4 * \width}
        \pgfmathsetmacro\xminval{#2 * \width}
        \pgfmathsetmacro\xmaxval{#2 * \width + \length}
        \pgfmathsetmacro\yminval{(1 - #3) * \height - \length}
        \pgfmathsetmacro\ymaxval{(1 - #3) * \height}

        \begin{scope}
            % Use fill with a rectangle, not a node
            \path[fill overzoom image = results/#1.jpg] (0, 0) rectangle (1, 1);
            % Now clip to viewport
            \clip (0, 0) rectangle (1, 1);
            \begin{scope}[xscale = 1, yscale = 1]
                \pgfmathsetmacro{\trimleft}{\xminval}
                \pgfmathsetmacro{\trimbottom}{\height - \ymaxval}  % Flip y-axis because trim starts from top-left
                \pgfmathsetmacro{\trimright}{\width - \xmaxval}
                \pgfmathsetmacro{\trimtop}{\yminval}
                
                \node at (0.5,0.5) {%
                    \includegraphics[
                        width = 1.78cm, height = 1.78cm,
                        trim = {\trimleft bp} {\trimbottom bp} {\trimright bp} {\trimtop bp},
                        clip
                    ]{results/#1.jpg}%
                };
            \end{scope}
        \end{scope}

        \begin{scope}
            \draw[#5, thick] (0, 0) rectangle (1, 1);
        \end{scope}
    \endgroup
}

\def\arraystretch{.5}
\setlength\tabcolsep{1pt}

\begin{tabularx}{\linewidth}{@{}cc@{}}
    \begin{minipage}[t]{.48\linewidth}
    \vspace{0pt}%
    \begin{tikzpicture}[scale = 5.5]
        \SplitThreeImages{\scene/\firstmethod}{\scene/\secondmethod}{\scene/\thirdmethod}
        \begin{scope}
            \draw[\firstcropcolor, ultra thick]
            (\aspectratio * \firstcropxmin, \firstcropymin)
            rectangle
            (\aspectratio * \firstcropxmin + \aspectratio * \croplength, \firstcropymin + \aspectratio * \croplength);
            
            \draw[\secondcropcolor, ultra thick]
            (\aspectratio * \secondcropxmin, \secondcropymin)
            rectangle
            (\aspectratio * \secondcropxmin + \aspectratio * \croplength, \secondcropymin + \aspectratio * \croplength);
        \end{scope}
        \begin{scope}
            \contourlength{.1em}
            \MethodLabelAt{.025}{.9}{west}{\firstmethoddisplaynamea}{\firstmethoddisplaynameb}
            \MethodLabelAt{.69}{.9}{west}{\secondmethoddisplaynamea}{\secondmethoddisplaynameb}
            \MethodLabelAt{1.3}{.9}{west}{\thirdmethoddisplaynamea}{\thirdmethoddisplaynameb}
        \end{scope}
    \end{tikzpicture}
    \end{minipage}
    &
    \hspace{1.25cm}
    \begin{minipage}[t]{.48\linewidth}
    \vspace{0pt}%
    \begin{tabular}{@{}ccccc@{}}
        % row 1
        \begin{tikzpicture}[scale = 1.78]
            \CropImage{\scene/\firstmethod}\firstcropxmin\firstcropymin\croplength{\firstcropcolor}
        \end{tikzpicture}
        &
        \begin{tikzpicture}[scale = 1.78]
            \CropImage{\scene/\firstmethod-MAPE}\firstcropxmin\firstcropymin\croplength{\firstcropcolor}
        \end{tikzpicture}
        &
        \begin{tikzpicture}[scale = 1.78]
            \CropImage{\scene/\firstmethod}\secondcropxmin\secondcropymin\croplength{\secondcropcolor}
        \end{tikzpicture}
        &
        \begin{tikzpicture}[scale = 1.78]
            \CropImage{\scene/\firstmethod-MAPE}\secondcropxmin\secondcropymin\croplength{\secondcropcolor}
        \end{tikzpicture}
        &
        \RotTableMethodLabel{\firstmethoddisplaynamea}{\firstmethoddisplaynameb}
        \\
        % row 2
        \begin{tikzpicture}[scale = 1.78]
            \CropImage{\scene/\secondmethod}\firstcropxmin\firstcropymin\croplength{\firstcropcolor}
        \end{tikzpicture}
        &
        \begin{tikzpicture}[scale = 1.78]
          \CropImage{\scene/\secondmethod-MAPE}\firstcropxmin\firstcropymin\croplength{\firstcropcolor}
        \end{tikzpicture}
        &
        \begin{tikzpicture}[scale = 1.78]
            \CropImage{\scene/\secondmethod}\secondcropxmin\secondcropymin\croplength{\secondcropcolor}
        \end{tikzpicture}
        &
        \begin{tikzpicture}[scale = 1.78]
            \CropImage{\scene/\secondmethod-MAPE}\secondcropxmin\secondcropymin\croplength{\secondcropcolor}
        \end{tikzpicture}
        &
        \RotTableMethodLabel{\secondmethoddisplaynamea}{\secondmethoddisplaynameb}
        \\
        % row 3
        \begin{tikzpicture}[scale = 1.78]
            \CropImage{\scene/\thirdmethod}\firstcropxmin\firstcropymin\croplength{\firstcropcolor}
        \end{tikzpicture}
        &
        \begin{tikzpicture}[scale = 1.78]
            \CropImage{\scene/\thirdmethod-MAPE}\firstcropxmin\firstcropymin\croplength{\firstcropcolor}
        \end{tikzpicture}
        &
        \begin{tikzpicture}[scale = 1.78]
            \CropImage{\scene/\thirdmethod}\secondcropxmin\secondcropymin\croplength{\secondcropcolor}
        \end{tikzpicture}
        &
        \begin{tikzpicture}[scale = 1.78]
            \CropImage{\scene/\thirdmethod-MAPE}\secondcropxmin\secondcropymin\croplength{\secondcropcolor}
        \end{tikzpicture}
        &
        \RotTableMethodLabel{\thirdmethoddisplaynamea}{\thirdmethoddisplaynameb}
    \end{tabular}
    \end{minipage}
\end{tabularx}

    \caption{Equal-rendering-time comparison ($1/30$ seconds of computation) of \gls{pt} (left), \gls{mala} Restore (middle), and Diffusion Restore (right) for the \textsc{Veach, Ajar} scene provided by \citet{bitterli2016resources}. The inset error maps show the corresponding \gls{mape} for each rendering.}
    \label{fig:veach-ajar}
\end{figure*}

\begin{figure*}
    %!TEX root = ../main.tex

% Rotated two-line method label (for right margin of table)
% #1 first line, #2 second line
\newcommand{\RotTableMethodLabel}[2]{%
    \smash{%
        \raisebox{0.5\height}{% optisch zentrieren; bei Bedarf feinjustieren
            \rotatebox[origin = c]{90}{%
                \tiny\begin{tabular}{@{}l@{}}
                    \hspace{.6cm}#1\\[.1ex]
                    \hspace{.6cm}#2
                \end{tabular}%
            }%
        }%
    }%
}

% #1 = x position
% #2 = y position
% #3 = anchor
% #3 = first line
% #4 = second line
\newcommand{\MethodLabelAt}[5]{%
  \node[anchor = #3] at (#1, #2) {%
    \shortstack[l]{%
      \contour{black}{\small\textcolor{white}{#4}}\\[-0.2ex]%
      \contour{black}{\small\textcolor{white}{#5}}%
    }%
  };%
}

% Two-line tiny label for tables (left-aligned)
\newcommand{\TableMethodLabel}[2]{%
  \begin{tabular}{@{}l@{}}%
    \tiny #1\\[-0.1ex]%
    \tiny #2%
  \end{tabular}%
}

% \pgfmathsetmacro\width{1024}
% \pgfmathsetmacro\height{768}
\pgfmathsetmacro\width{1024}
\pgfmathsetmacro\height{576}
\pgfmathsetmacro\aspectratio{16 / 9}

\newcommand\scene{torus}

\newcommand\firstmethod{Path_Tracing}
\newcommand\secondmethod{MALA_Restore}
\newcommand\thirdmethod{Diffusion_Restore}
\newcommand\firstmethoddisplaynamea{Path Tracing}
\newcommand\firstmethoddisplaynameb{\vphantom|}
\newcommand\secondmethoddisplaynamea{MALA Restore}
\newcommand\secondmethoddisplaynameb{\vphantom|}
\newcommand\thirdmethoddisplaynamea{Diffusion Restore}
\newcommand\thirdmethoddisplaynameb{\vphantom|}

\pgfmathsetmacro\croplength{.14}
\pgfmathsetmacro\firstcropxmin{.18}
\pgfmathsetmacro\firstcropymin{.28}
\pgfmathsetmacro\secondcropxmin{.65}
\pgfmathsetmacro\secondcropymin{.21}
\newcommand\firstcropcolor{red}
\newcommand\secondcropcolor{blue}

\newcommand{\SplitThreeImages}[3]{
    \begin{scope}
       \clip (0, 0) -- (.2 * \aspectratio, 0) -- (.3 * \aspectratio, 1) -- (0, 1) -- cycle;
        \path[fill overzoom image = results/#1.jpg] (0, 0) rectangle (\aspectratio, 1);
    \end{scope}
    \begin{scope}
       \clip (.2 * \aspectratio, 0) -- (0.6 * \aspectratio, 0) -- (0.7 * \aspectratio, 1) -- (0.3 * \aspectratio, 1) -- cycle;
        \path[fill overzoom image = results/#2.jpg] (0, 0) rectangle (\aspectratio, 1);
    \end{scope}
   \begin{scope}
       \clip (.6 * \aspectratio, 0) -- (\aspectratio, 0) -- (\aspectratio, 1) -- (0.7 * \aspectratio, 1) -- cycle;
       \path[fill overzoom image = results/#3.jpg] (0,0) rectangle (\aspectratio, 1);
   \end{scope}
   \begin{scope}
        \draw[black, thick] (0, 0) rectangle (\aspectratio, 1);
        \draw[draw=black,thick] (.2 * \aspectratio, 0) -- (.3 * \aspectratio, 1);
        \draw[draw=black,thick] (.6 * \aspectratio, 0) -- (.7 * \aspectratio, 1);
   \end{scope}
}

%   - #1: scene name (without extension; image is results/#1.jpg)
%   - #2: crop x (normalized, 0–1; 0=left)
%   - #3: crop y (normalized, 0–1; 0=bottom; note: image origin is lower left)
%   - #4: crop length (normalized)
%   - #5: color (for drawing the outline)
\newcommand{\CropImage}[5]{%
    \begingroup
        % Convert normalized crop to pixel values
        \pgfmathsetmacro\length{#4 * \width}
        \pgfmathsetmacro\xminval{#2 * \width}
        \pgfmathsetmacro\xmaxval{#2 * \width + \length}
        \pgfmathsetmacro\yminval{(1 - #3) * \height - \length}
        \pgfmathsetmacro\ymaxval{(1 - #3) * \height}

        \begin{scope}
            % Use fill with a rectangle, not a node
            \path[fill overzoom image = results/#1.jpg] (0, 0) rectangle (1, 1);
            % Now clip to viewport
            \clip (0, 0) rectangle (1, 1);
            \begin{scope}[xscale = 1, yscale = 1]
                \pgfmathsetmacro{\trimleft}{\xminval}
                \pgfmathsetmacro{\trimbottom}{\height - \ymaxval}  % Flip y-axis because trim starts from top-left
                \pgfmathsetmacro{\trimright}{\width - \xmaxval}
                \pgfmathsetmacro{\trimtop}{\yminval}
                
                \node at (0.5,0.5) {%
                    \includegraphics[
                        width = 1.78cm, height = 1.78cm,
                        trim = {\trimleft bp} {\trimbottom bp} {\trimright bp} {\trimtop bp},
                        clip
                    ]{results/#1.jpg}%
                };
            \end{scope}
        \end{scope}

        \begin{scope}
            \draw[#5, thick] (0, 0) rectangle (1, 1);
        \end{scope}
    \endgroup
}

\def\arraystretch{.5}
\setlength\tabcolsep{1pt}

\begin{tabularx}{\linewidth}{@{}cc@{}}
    \begin{minipage}[t]{.48\linewidth}
    \vspace{0pt}%
    \begin{tikzpicture}[scale = 5.5]
        \SplitThreeImages{\scene/\firstmethod}{\scene/\secondmethod}{\scene/\thirdmethod}
        \begin{scope}
            \draw[\firstcropcolor, ultra thick]
            (\aspectratio * \firstcropxmin, \firstcropymin)
            rectangle
            (\aspectratio * \firstcropxmin + \aspectratio * \croplength, \firstcropymin + \aspectratio * \croplength);
            
            \draw[\secondcropcolor, ultra thick]
            (\aspectratio * \secondcropxmin, \secondcropymin)
            rectangle
            (\aspectratio * \secondcropxmin + \aspectratio * \croplength, \secondcropymin + \aspectratio * \croplength);
        \end{scope}
        \begin{scope}
            \contourlength{.1em}
            \MethodLabelAt{.025}{.9}{west}{\firstmethoddisplaynamea}{\firstmethoddisplaynameb}
            \MethodLabelAt{.69}{.9}{west}{\secondmethoddisplaynamea}{\secondmethoddisplaynameb}
            \MethodLabelAt{1.3}{.9}{west}{\thirdmethoddisplaynamea}{\thirdmethoddisplaynameb}
        \end{scope}
    \end{tikzpicture}
    \end{minipage}
    &
    \hspace{1.25cm}
    \begin{minipage}[t]{.48\linewidth}
    \vspace{0pt}%
    \begin{tabular}{@{}ccccc@{}}
        % row 1
        \begin{tikzpicture}[scale = 1.78]
            \CropImage{\scene/\firstmethod}\firstcropxmin\firstcropymin\croplength{\firstcropcolor}
        \end{tikzpicture}
        &
        \begin{tikzpicture}[scale = 1.78]
            \CropImage{\scene/\firstmethod-MAPE}\firstcropxmin\firstcropymin\croplength{\firstcropcolor}
        \end{tikzpicture}
        &
        \begin{tikzpicture}[scale = 1.78]
            \CropImage{\scene/\firstmethod}\secondcropxmin\secondcropymin\croplength{\secondcropcolor}
        \end{tikzpicture}
        &
        \begin{tikzpicture}[scale = 1.78]
            \CropImage{\scene/\firstmethod-MAPE}\secondcropxmin\secondcropymin\croplength{\secondcropcolor}
        \end{tikzpicture}
        &
        \RotTableMethodLabel{\firstmethoddisplaynamea}{\firstmethoddisplaynameb}
        \\
        % row 2
        \begin{tikzpicture}[scale = 1.78]
            \CropImage{\scene/\secondmethod}\firstcropxmin\firstcropymin\croplength{\firstcropcolor}
        \end{tikzpicture}
        &
        \begin{tikzpicture}[scale = 1.78]
          \CropImage{\scene/\secondmethod-MAPE}\firstcropxmin\firstcropymin\croplength{\firstcropcolor}
        \end{tikzpicture}
        &
        \begin{tikzpicture}[scale = 1.78]
            \CropImage{\scene/\secondmethod}\secondcropxmin\secondcropymin\croplength{\secondcropcolor}
        \end{tikzpicture}
        &
        \begin{tikzpicture}[scale = 1.78]
            \CropImage{\scene/\secondmethod-MAPE}\secondcropxmin\secondcropymin\croplength{\secondcropcolor}
        \end{tikzpicture}
        &
        \RotTableMethodLabel{\secondmethoddisplaynamea}{\secondmethoddisplaynameb}
        \\
        % row 3
        \begin{tikzpicture}[scale = 1.78]
            \CropImage{\scene/\thirdmethod}\firstcropxmin\firstcropymin\croplength{\firstcropcolor}
        \end{tikzpicture}
        &
        \begin{tikzpicture}[scale = 1.78]
            \CropImage{\scene/\thirdmethod-MAPE}\firstcropxmin\firstcropymin\croplength{\firstcropcolor}
        \end{tikzpicture}
        &
        \begin{tikzpicture}[scale = 1.78]
            \CropImage{\scene/\thirdmethod}\secondcropxmin\secondcropymin\croplength{\secondcropcolor}
        \end{tikzpicture}
        &
        \begin{tikzpicture}[scale = 1.78]
            \CropImage{\scene/\thirdmethod-MAPE}\secondcropxmin\secondcropymin\croplength{\secondcropcolor}
        \end{tikzpicture}
        &
        \RotTableMethodLabel{\thirdmethoddisplaynamea}{\thirdmethoddisplaynameb}
    \end{tabular}
    \end{minipage}
\end{tabularx}

    \caption{Equal-rendering-time comparison ($1/30$ seconds of computation) of \gls{pt} (left), \gls{mala} Restore (middle), and Diffusion Restore (right) for the \textsc{Torus} scene provided by \citet{li2015anisotropic}. The inset error maps show the corresponding \gls{mape} for each rendering.}
    \label{fig:torus}
\end{figure*}

\begin{figure*}
    %!TEX root = ../main.tex

% Rotated two-line method label (for right margin of table)
% #1 first line, #2 second line
\newcommand{\RotTableMethodLabel}[2]{%
    \smash{%
        \raisebox{0.5\height}{% optisch zentrieren; bei Bedarf feinjustieren
            \rotatebox[origin = c]{90}{%
                \tiny\begin{tabular}{@{}l@{}}
                    \hspace{.6cm}#1\\[.1ex]
                    \hspace{.6cm}#2
                \end{tabular}%
            }%
        }%
    }%
}

% #1 = x position
% #2 = y position
% #3 = anchor
% #3 = first line
% #4 = second line
\newcommand{\MethodLabelAt}[5]{%
  \node[anchor = #3] at (#1, #2) {%
    \shortstack[l]{%
      \contour{black}{\small\textcolor{white}{#4}}\\[-0.2ex]%
      \contour{black}{\small\textcolor{white}{#5}}%
    }%
  };%
}

% Two-line tiny label for tables (left-aligned)
\newcommand{\TableMethodLabel}[2]{%
  \begin{tabular}{@{}l@{}}%
    \tiny #1\\[-0.1ex]%
    \tiny #2%
  \end{tabular}%
}

% \pgfmathsetmacro\width{1024}
% \pgfmathsetmacro\height{768}
\pgfmathsetmacro\width{1024}
\pgfmathsetmacro\height{576}
\pgfmathsetmacro\aspectratio{16 / 9}

\newcommand\scene{glass_of_water}

\newcommand\firstmethod{Path_Tracing}
\newcommand\secondmethod{MALA_Restore}
\newcommand\thirdmethod{Diffusion_Restore}
\newcommand\firstmethoddisplaynamea{Path Tracing}
\newcommand\firstmethoddisplaynameb{\vphantom|}
\newcommand\secondmethoddisplaynamea{MALA Restore}
\newcommand\secondmethoddisplaynameb{\vphantom|}
\newcommand\thirdmethoddisplaynamea{Diffusion Restore}
\newcommand\thirdmethoddisplaynameb{\vphantom|}

\pgfmathsetmacro\croplength{.12}
\pgfmathsetmacro\firstcropxmin{.42}
\pgfmathsetmacro\firstcropymin{.77}
\pgfmathsetmacro\secondcropxmin{.45}
\pgfmathsetmacro\secondcropymin{.25}
\newcommand\firstcropcolor{red}
\newcommand\secondcropcolor{blue}

\newcommand{\SplitThreeImages}[3]{
    \begin{scope}
       \clip (0, 0) -- (.2 * \aspectratio, 0) -- (.3 * \aspectratio, 1) -- (0, 1) -- cycle;
        \path[fill overzoom image = results/#1.jpg] (0, 0) rectangle (\aspectratio, 1);
    \end{scope}
    \begin{scope}
       \clip (.2 * \aspectratio, 0) -- (0.6 * \aspectratio, 0) -- (0.7 * \aspectratio, 1) -- (0.3 * \aspectratio, 1) -- cycle;
        \path[fill overzoom image = results/#2.jpg] (0, 0) rectangle (\aspectratio, 1);
    \end{scope}
   \begin{scope}
       \clip (.6 * \aspectratio, 0) -- (\aspectratio, 0) -- (\aspectratio, 1) -- (0.7 * \aspectratio, 1) -- cycle;
       \path[fill overzoom image = results/#3.jpg] (0,0) rectangle (\aspectratio, 1);
   \end{scope}
   \begin{scope}
        \draw[black, thick] (0, 0) rectangle (\aspectratio, 1);
        \draw[draw=black,thick] (.2 * \aspectratio, 0) -- (.3 * \aspectratio, 1);
        \draw[draw=black,thick] (.6 * \aspectratio, 0) -- (.7 * \aspectratio, 1);
   \end{scope}
}

%   - #1: scene name (without extension; image is results/#1.jpg)
%   - #2: crop x (normalized, 0–1; 0=left)
%   - #3: crop y (normalized, 0–1; 0=bottom; note: image origin is lower left)
%   - #4: crop length (normalized)
%   - #5: color (for drawing the outline)
\newcommand{\CropImage}[5]{%
    \begingroup
        % Convert normalized crop to pixel values
        \pgfmathsetmacro\length{#4 * \width}
        \pgfmathsetmacro\xminval{#2 * \width}
        \pgfmathsetmacro\xmaxval{#2 * \width + \length}
        \pgfmathsetmacro\yminval{(1 - #3) * \height - \length}
        \pgfmathsetmacro\ymaxval{(1 - #3) * \height}

        \begin{scope}
            % Use fill with a rectangle, not a node
            \path[fill overzoom image = results/#1.jpg] (0, 0) rectangle (1, 1);
            % Now clip to viewport
            \clip (0, 0) rectangle (1, 1);
            \begin{scope}[xscale = 1, yscale = 1]
                \pgfmathsetmacro{\trimleft}{\xminval}
                \pgfmathsetmacro{\trimbottom}{\height - \ymaxval}  % Flip y-axis because trim starts from top-left
                \pgfmathsetmacro{\trimright}{\width - \xmaxval}
                \pgfmathsetmacro{\trimtop}{\yminval}
                
                \node at (0.5,0.5) {%
                    \includegraphics[
                        width = 1.78cm, height = 1.78cm,
                        trim = {\trimleft bp} {\trimbottom bp} {\trimright bp} {\trimtop bp},
                        clip
                    ]{results/#1.jpg}%
                };
            \end{scope}
        \end{scope}

        \begin{scope}
            \draw[#5, thick] (0, 0) rectangle (1, 1);
        \end{scope}
    \endgroup
}

\def\arraystretch{.5}
\setlength\tabcolsep{1pt}

\begin{tabularx}{\linewidth}{@{}cc@{}}
    \begin{minipage}[t]{.48\linewidth}
    \vspace{0pt}%
    \begin{tikzpicture}[scale = 5.5]
        \SplitThreeImages{\scene/\firstmethod}{\scene/\secondmethod}{\scene/\thirdmethod}
        \begin{scope}
            \draw[\firstcropcolor, ultra thick]
            (\aspectratio * \firstcropxmin, \firstcropymin)
            rectangle
            (\aspectratio * \firstcropxmin + \aspectratio * \croplength, \firstcropymin + \aspectratio * \croplength);
            
            \draw[\secondcropcolor, ultra thick]
            (\aspectratio * \secondcropxmin, \secondcropymin)
            rectangle
            (\aspectratio * \secondcropxmin + \aspectratio * \croplength, \secondcropymin + \aspectratio * \croplength);
        \end{scope}
        \begin{scope}
            \contourlength{.1em}
            \MethodLabelAt{.025}{.9}{west}{\firstmethoddisplaynamea}{\firstmethoddisplaynameb}
            \MethodLabelAt{.69}{.9}{west}{\secondmethoddisplaynamea}{\secondmethoddisplaynameb}
            \MethodLabelAt{1.3}{.9}{west}{\thirdmethoddisplaynamea}{\thirdmethoddisplaynameb}
        \end{scope}
    \end{tikzpicture}
    \end{minipage}
    &
    \hspace{1.25cm}
    \begin{minipage}[t]{.48\linewidth}
    \vspace{0pt}%
    \begin{tabular}{@{}ccccc@{}}
        % row 1
        \begin{tikzpicture}[scale = 1.78]
            \CropImage{\scene/\firstmethod}\firstcropxmin\firstcropymin\croplength{\firstcropcolor}
        \end{tikzpicture}
        &
        \begin{tikzpicture}[scale = 1.78]
            \CropImage{\scene/\firstmethod-MAPE}\firstcropxmin\firstcropymin\croplength{\firstcropcolor}
        \end{tikzpicture}
        &
        \begin{tikzpicture}[scale = 1.78]
            \CropImage{\scene/\firstmethod}\secondcropxmin\secondcropymin\croplength{\secondcropcolor}
        \end{tikzpicture}
        &
        \begin{tikzpicture}[scale = 1.78]
            \CropImage{\scene/\firstmethod-MAPE}\secondcropxmin\secondcropymin\croplength{\secondcropcolor}
        \end{tikzpicture}
        &
        \RotTableMethodLabel{\firstmethoddisplaynamea}{\firstmethoddisplaynameb}
        \\
        % row 2
        \begin{tikzpicture}[scale = 1.78]
            \CropImage{\scene/\secondmethod}\firstcropxmin\firstcropymin\croplength{\firstcropcolor}
        \end{tikzpicture}
        &
        \begin{tikzpicture}[scale = 1.78]
          \CropImage{\scene/\secondmethod-MAPE}\firstcropxmin\firstcropymin\croplength{\firstcropcolor}
        \end{tikzpicture}
        &
        \begin{tikzpicture}[scale = 1.78]
            \CropImage{\scene/\secondmethod}\secondcropxmin\secondcropymin\croplength{\secondcropcolor}
        \end{tikzpicture}
        &
        \begin{tikzpicture}[scale = 1.78]
            \CropImage{\scene/\secondmethod-MAPE}\secondcropxmin\secondcropymin\croplength{\secondcropcolor}
        \end{tikzpicture}
        &
        \RotTableMethodLabel{\secondmethoddisplaynamea}{\secondmethoddisplaynameb}
        \\
        % row 3
        \begin{tikzpicture}[scale = 1.78]
            \CropImage{\scene/\thirdmethod}\firstcropxmin\firstcropymin\croplength{\firstcropcolor}
        \end{tikzpicture}
        &
        \begin{tikzpicture}[scale = 1.78]
            \CropImage{\scene/\thirdmethod-MAPE}\firstcropxmin\firstcropymin\croplength{\firstcropcolor}
        \end{tikzpicture}
        &
        \begin{tikzpicture}[scale = 1.78]
            \CropImage{\scene/\thirdmethod}\secondcropxmin\secondcropymin\croplength{\secondcropcolor}
        \end{tikzpicture}
        &
        \begin{tikzpicture}[scale = 1.78]
            \CropImage{\scene/\thirdmethod-MAPE}\secondcropxmin\secondcropymin\croplength{\secondcropcolor}
        \end{tikzpicture}
        &
        \RotTableMethodLabel{\thirdmethoddisplaynamea}{\thirdmethoddisplaynameb}
    \end{tabular}
    \end{minipage}
\end{tabularx}

    \caption{Equal-rendering-time comparison ($1/30$ seconds of computation) of \gls{pt} (left), \gls{mala} Restore (middle), and Diffusion Restore (right) for the \textsc{Glass of Water} scene provided by \citet{bitterli2016resources}. The inset error maps show the corresponding \gls{mape} for each rendering.}
    \label{fig:glass-of-water}
\end{figure*}

\begin{figure*}
    %!TEX root = ../main.tex

% Rotated two-line method label (for right margin of table)
% #1 first line, #2 second line
\newcommand{\RotTableMethodLabel}[2]{%
    \smash{%
        \raisebox{0.5\height}{% optisch zentrieren; bei Bedarf feinjustieren
            \rotatebox[origin = c]{90}{%
                \tiny\begin{tabular}{@{}l@{}}
                    \hspace{.6cm}#1\\[.1ex]
                    \hspace{.6cm}#2
                \end{tabular}%
            }%
        }%
    }%
}

% #1 = x position
% #2 = y position
% #3 = anchor
% #3 = first line
% #4 = second line
\newcommand{\MethodLabelAt}[5]{%
  \node[anchor = #3] at (#1, #2) {%
    \shortstack[l]{%
      \contour{black}{\small\textcolor{white}{#4}}\\[-0.2ex]%
      \contour{black}{\small\textcolor{white}{#5}}%
    }%
  };%
}

% Two-line tiny label for tables (left-aligned)
\newcommand{\TableMethodLabel}[2]{%
  \begin{tabular}{@{}l@{}}%
    \tiny #1\\[-0.1ex]%
    \tiny #2%
  \end{tabular}%
}

% \pgfmathsetmacro\width{1024}
% \pgfmathsetmacro\height{768}
\pgfmathsetmacro\width{1024}
\pgfmathsetmacro\height{576}
\pgfmathsetmacro\aspectratio{16 / 9}

\newcommand\scene{bathroom}

\newcommand\firstmethod{Path_Tracing}
\newcommand\secondmethod{MALA_Restore}
\newcommand\thirdmethod{Diffusion_Restore}
\newcommand\firstmethoddisplaynamea{Path Tracing}
\newcommand\firstmethoddisplaynameb{\vphantom|}
\newcommand\secondmethoddisplaynamea{MALA Restore}
\newcommand\secondmethoddisplaynameb{\vphantom|}
\newcommand\thirdmethoddisplaynamea{Diffusion Restore}
\newcommand\thirdmethoddisplaynameb{\vphantom|}

\pgfmathsetmacro\croplength{.12}
\pgfmathsetmacro\firstcropxmin{.1}
\pgfmathsetmacro\firstcropymin{.75}
\pgfmathsetmacro\secondcropxmin{.12}
\pgfmathsetmacro\secondcropymin{.138}
\newcommand\firstcropcolor{red}
\newcommand\secondcropcolor{blue}

\newcommand{\SplitThreeImages}[3]{
    \begin{scope}
       \clip (0, 0) -- (.2 * \aspectratio, 0) -- (.3 * \aspectratio, 1) -- (0, 1) -- cycle;
        \path[fill overzoom image = results/#1.jpg] (0, 0) rectangle (\aspectratio, 1);
    \end{scope}
    \begin{scope}
       \clip (.2 * \aspectratio, 0) -- (0.6 * \aspectratio, 0) -- (0.7 * \aspectratio, 1) -- (0.3 * \aspectratio, 1) -- cycle;
        \path[fill overzoom image = results/#2.jpg] (0, 0) rectangle (\aspectratio, 1);
    \end{scope}
   \begin{scope}
       \clip (.6 * \aspectratio, 0) -- (\aspectratio, 0) -- (\aspectratio, 1) -- (0.7 * \aspectratio, 1) -- cycle;
       \path[fill overzoom image = results/#3.jpg] (0,0) rectangle (\aspectratio, 1);
   \end{scope}
   \begin{scope}
        \draw[black, thick] (0, 0) rectangle (\aspectratio, 1);
        \draw[draw=black,thick] (.2 * \aspectratio, 0) -- (.3 * \aspectratio, 1);
        \draw[draw=black,thick] (.6 * \aspectratio, 0) -- (.7 * \aspectratio, 1);
   \end{scope}
}

%   - #1: scene name (without extension; image is results/#1.jpg)
%   - #2: crop x (normalized, 0–1; 0=left)
%   - #3: crop y (normalized, 0–1; 0=bottom; note: image origin is lower left)
%   - #4: crop length (normalized)
%   - #5: color (for drawing the outline)
\newcommand{\CropImage}[5]{%
    \begingroup
        % Convert normalized crop to pixel values
        \pgfmathsetmacro\length{#4 * \width}
        \pgfmathsetmacro\xminval{#2 * \width}
        \pgfmathsetmacro\xmaxval{#2 * \width + \length}
        \pgfmathsetmacro\yminval{(1 - #3) * \height - \length}
        \pgfmathsetmacro\ymaxval{(1 - #3) * \height}

        \begin{scope}
            % Use fill with a rectangle, not a node
            \path[fill overzoom image = results/#1.jpg] (0, 0) rectangle (1, 1);
            % Now clip to viewport
            \clip (0, 0) rectangle (1, 1);
            \begin{scope}[xscale = 1, yscale = 1]
                \pgfmathsetmacro{\trimleft}{\xminval}
                \pgfmathsetmacro{\trimbottom}{\height - \ymaxval}  % Flip y-axis because trim starts from top-left
                \pgfmathsetmacro{\trimright}{\width - \xmaxval}
                \pgfmathsetmacro{\trimtop}{\yminval}
                
                \node at (0.5,0.5) {%
                    \includegraphics[
                        width = 1.78cm, height = 1.78cm,
                        trim = {\trimleft bp} {\trimbottom bp} {\trimright bp} {\trimtop bp},
                        clip
                    ]{results/#1.jpg}%
                };
            \end{scope}
        \end{scope}

        \begin{scope}
            \draw[#5, thick] (0, 0) rectangle (1, 1);
        \end{scope}
    \endgroup
}

\def\arraystretch{.5}
\setlength\tabcolsep{1pt}

\begin{tabularx}{\linewidth}{@{}cc@{}}
    \begin{minipage}[t]{.48\linewidth}
    \vspace{0pt}%
    \begin{tikzpicture}[scale = 5.5]
        \SplitThreeImages{\scene/\firstmethod}{\scene/\secondmethod}{\scene/\thirdmethod}
        \begin{scope}
            \draw[\firstcropcolor, ultra thick]
            (\aspectratio * \firstcropxmin, \firstcropymin)
            rectangle
            (\aspectratio * \firstcropxmin + \aspectratio * \croplength, \firstcropymin + \aspectratio * \croplength);
            
            \draw[\secondcropcolor, ultra thick]
            (\aspectratio * \secondcropxmin, \secondcropymin)
            rectangle
            (\aspectratio * \secondcropxmin + \aspectratio * \croplength, \secondcropymin + \aspectratio * \croplength);
        \end{scope}
        \begin{scope}
            \contourlength{.1em}
            \MethodLabelAt{.025}{.9}{west}{\firstmethoddisplaynamea}{\firstmethoddisplaynameb}
            \MethodLabelAt{.69}{.9}{west}{\secondmethoddisplaynamea}{\secondmethoddisplaynameb}
            \MethodLabelAt{1.3}{.9}{west}{\thirdmethoddisplaynamea}{\thirdmethoddisplaynameb}
        \end{scope}
    \end{tikzpicture}
    \end{minipage}
    &
    \hspace{1.25cm}
    \begin{minipage}[t]{.48\linewidth}
    \vspace{0pt}%
    \begin{tabular}{@{}ccccc@{}}
        % row 1
        \begin{tikzpicture}[scale = 1.78]
            \CropImage{\scene/\firstmethod}\firstcropxmin\firstcropymin\croplength{\firstcropcolor}
        \end{tikzpicture}
        &
        \begin{tikzpicture}[scale = 1.78]
            \CropImage{\scene/\firstmethod-MAPE}\firstcropxmin\firstcropymin\croplength{\firstcropcolor}
        \end{tikzpicture}
        &
        \begin{tikzpicture}[scale = 1.78]
            \CropImage{\scene/\firstmethod}\secondcropxmin\secondcropymin\croplength{\secondcropcolor}
        \end{tikzpicture}
        &
        \begin{tikzpicture}[scale = 1.78]
            \CropImage{\scene/\firstmethod-MAPE}\secondcropxmin\secondcropymin\croplength{\secondcropcolor}
        \end{tikzpicture}
        &
        \RotTableMethodLabel{\firstmethoddisplaynamea}{\firstmethoddisplaynameb}
        \\
        % row 2
        \begin{tikzpicture}[scale = 1.78]
            \CropImage{\scene/\secondmethod}\firstcropxmin\firstcropymin\croplength{\firstcropcolor}
        \end{tikzpicture}
        &
        \begin{tikzpicture}[scale = 1.78]
          \CropImage{\scene/\secondmethod-MAPE}\firstcropxmin\firstcropymin\croplength{\firstcropcolor}
        \end{tikzpicture}
        &
        \begin{tikzpicture}[scale = 1.78]
            \CropImage{\scene/\secondmethod}\secondcropxmin\secondcropymin\croplength{\secondcropcolor}
        \end{tikzpicture}
        &
        \begin{tikzpicture}[scale = 1.78]
            \CropImage{\scene/\secondmethod-MAPE}\secondcropxmin\secondcropymin\croplength{\secondcropcolor}
        \end{tikzpicture}
        &
        \RotTableMethodLabel{\secondmethoddisplaynamea}{\secondmethoddisplaynameb}
        \\
        % row 3
        \begin{tikzpicture}[scale = 1.78]
            \CropImage{\scene/\thirdmethod}\firstcropxmin\firstcropymin\croplength{\firstcropcolor}
        \end{tikzpicture}
        &
        \begin{tikzpicture}[scale = 1.78]
            \CropImage{\scene/\thirdmethod-MAPE}\firstcropxmin\firstcropymin\croplength{\firstcropcolor}
        \end{tikzpicture}
        &
        \begin{tikzpicture}[scale = 1.78]
            \CropImage{\scene/\thirdmethod}\secondcropxmin\secondcropymin\croplength{\secondcropcolor}
        \end{tikzpicture}
        &
        \begin{tikzpicture}[scale = 1.78]
            \CropImage{\scene/\thirdmethod-MAPE}\secondcropxmin\secondcropymin\croplength{\secondcropcolor}
        \end{tikzpicture}
        &
        \RotTableMethodLabel{\thirdmethoddisplaynamea}{\thirdmethoddisplaynameb}
    \end{tabular}
    \end{minipage}
\end{tabularx}

    \caption{Equal-rendering-time comparison ($1/30$ seconds of computation) of \gls{pt} (left), \gls{mala} Restore (middle), and Diffusion Restore (right) for the \textsc{Salle de Bain} scene provided by \citet{bitterli2016resources}. The inset error maps show the corresponding \gls{mape} for each rendering.}
    \label{fig:bathroom}
\end{figure*}

\begin{figure*}
    %!TEX root = ../main.tex

% Rotated two-line method label (for right margin of table)
% #1 first line, #2 second line
\newcommand{\RotTableMethodLabel}[2]{%
    \smash{%
        \raisebox{0.5\height}{% optisch zentrieren; bei Bedarf feinjustieren
            \rotatebox[origin = c]{90}{%
                \tiny\begin{tabular}{@{}l@{}}
                    \hspace{.6cm}#1\\[.1ex]
                    \hspace{.6cm}#2
                \end{tabular}%
            }%
        }%
    }%
}

% #1 = x position
% #2 = y position
% #3 = anchor
% #3 = first line
% #4 = second line
\newcommand{\MethodLabelAt}[5]{%
  \node[anchor = #3] at (#1, #2) {%
    \shortstack[l]{%
      \contour{black}{\small\textcolor{white}{#4}}\\[-0.2ex]%
      \contour{black}{\small\textcolor{white}{#5}}%
    }%
  };%
}

% Two-line tiny label for tables (left-aligned)
\newcommand{\TableMethodLabel}[2]{%
  \begin{tabular}{@{}l@{}}%
    \tiny #1\\[-0.1ex]%
    \tiny #2%
  \end{tabular}%
}

% \pgfmathsetmacro\width{1024}
% \pgfmathsetmacro\height{768}
\pgfmathsetmacro\width{800}
\pgfmathsetmacro\height{450}
\pgfmathsetmacro\aspectratio{16 / 9}

\newcommand\scene{pool}

\newcommand\firstmethod{Path_Tracing}
\newcommand\secondmethod{MALA_Restore}
\newcommand\thirdmethod{Diffusion_Restore}
\newcommand\firstmethoddisplaynamea{Path Tracing}
\newcommand\firstmethoddisplaynameb{\vphantom|}
\newcommand\secondmethoddisplaynamea{MALA Restore}
\newcommand\secondmethoddisplaynameb{\vphantom|}
\newcommand\thirdmethoddisplaynamea{Diffusion Restore}
\newcommand\thirdmethoddisplaynameb{\vphantom|}

\pgfmathsetmacro\croplength{.12}
\pgfmathsetmacro\firstcropxmin{.13}
\pgfmathsetmacro\firstcropymin{.0225}
\pgfmathsetmacro\secondcropxmin{.65}
\pgfmathsetmacro\secondcropymin{.6}
\newcommand\firstcropcolor{red}
\newcommand\secondcropcolor{blue}

\newcommand{\SplitThreeImages}[3]{
    \begin{scope}
       \clip (0, 0) -- (.2 * \aspectratio, 0) -- (.3 * \aspectratio, 1) -- (0, 1) -- cycle;
        \path[fill overzoom image = results/#1.jpg] (0, 0) rectangle (\aspectratio, 1);
    \end{scope}
    \begin{scope}
       \clip (.2 * \aspectratio, 0) -- (0.6 * \aspectratio, 0) -- (0.7 * \aspectratio, 1) -- (0.3 * \aspectratio, 1) -- cycle;
        \path[fill overzoom image = results/#2.jpg] (0, 0) rectangle (\aspectratio, 1);
    \end{scope}
   \begin{scope}
       \clip (.6 * \aspectratio, 0) -- (\aspectratio, 0) -- (\aspectratio, 1) -- (0.7 * \aspectratio, 1) -- cycle;
       \path[fill overzoom image = results/#3.jpg] (0,0) rectangle (\aspectratio, 1);
   \end{scope}
   \begin{scope}
        \draw[black, thick] (0, 0) rectangle (\aspectratio, 1);
        \draw[draw=black,thick] (.2 * \aspectratio, 0) -- (.3 * \aspectratio, 1);
        \draw[draw=black,thick] (.6 * \aspectratio, 0) -- (.7 * \aspectratio, 1);
   \end{scope}
}

%   - #1: scene name (without extension; image is results/#1.jpg)
%   - #2: crop x (normalized, 0–1; 0=left)
%   - #3: crop y (normalized, 0–1; 0=bottom; note: image origin is lower left)
%   - #4: crop length (normalized)
%   - #5: color (for drawing the outline)
\newcommand{\CropImage}[5]{%
    \begingroup
        % Convert normalized crop to pixel values
        \pgfmathsetmacro\length{#4 * \width}
        \pgfmathsetmacro\xminval{#2 * \width}
        \pgfmathsetmacro\xmaxval{#2 * \width + \length}
        \pgfmathsetmacro\yminval{(1 - #3) * \height - \length}
        \pgfmathsetmacro\ymaxval{(1 - #3) * \height}

        \begin{scope}
            % Use fill with a rectangle, not a node
            \path[fill overzoom image = results/#1.jpg] (0, 0) rectangle (1, 1);
            % Now clip to viewport
            \clip (0, 0) rectangle (1, 1);
            \begin{scope}[xscale = 1, yscale = 1]
                \pgfmathsetmacro{\trimleft}{\xminval}
                \pgfmathsetmacro{\trimbottom}{\height - \ymaxval}  % Flip y-axis because trim starts from top-left
                \pgfmathsetmacro{\trimright}{\width - \xmaxval}
                \pgfmathsetmacro{\trimtop}{\yminval}
                
                \node at (0.5,0.5) {%
                    \includegraphics[
                        width = 1.78cm, height = 1.78cm,
                        trim = {\trimleft bp} {\trimbottom bp} {\trimright bp} {\trimtop bp},
                        clip
                    ]{results/#1.jpg}%
                };
            \end{scope}
        \end{scope}

        \begin{scope}
            \draw[#5, thick] (0, 0) rectangle (1, 1);
        \end{scope}
    \endgroup
}

\def\arraystretch{.5}
\setlength\tabcolsep{1pt}

\begin{tabularx}{\linewidth}{@{}cc@{}}
    \begin{minipage}[t]{.48\linewidth}
    \vspace{0pt}%
    \begin{tikzpicture}[scale = 5.5]
        \SplitThreeImages{\scene/\firstmethod}{\scene/\secondmethod}{\scene/\thirdmethod}
        \begin{scope}
            \draw[\firstcropcolor, ultra thick]
            (\aspectratio * \firstcropxmin, \firstcropymin)
            rectangle
            (\aspectratio * \firstcropxmin + \aspectratio * \croplength, \firstcropymin + \aspectratio * \croplength);
            
            \draw[\secondcropcolor, ultra thick]
            (\aspectratio * \secondcropxmin, \secondcropymin)
            rectangle
            (\aspectratio * \secondcropxmin + \aspectratio * \croplength, \secondcropymin + \aspectratio * \croplength);
        \end{scope}
        \begin{scope}
            \contourlength{.1em}
            \MethodLabelAt{.025}{.9}{west}{\firstmethoddisplaynamea}{\firstmethoddisplaynameb}
            \MethodLabelAt{.69}{.9}{west}{\secondmethoddisplaynamea}{\secondmethoddisplaynameb}
            \MethodLabelAt{1.3}{.9}{west}{\thirdmethoddisplaynamea}{\thirdmethoddisplaynameb}
        \end{scope}
    \end{tikzpicture}
    \end{minipage}
    &
    \hspace{1.25cm}
    \begin{minipage}[t]{.48\linewidth}
    \vspace{0pt}%
    \begin{tabular}{@{}ccccc@{}}
        % row 1
        \begin{tikzpicture}[scale = 1.78]
            \CropImage{\scene/\firstmethod}\firstcropxmin\firstcropymin\croplength{\firstcropcolor}
        \end{tikzpicture}
        &
        \begin{tikzpicture}[scale = 1.78]
            \CropImage{\scene/\firstmethod-MAPE}\firstcropxmin\firstcropymin\croplength{\firstcropcolor}
        \end{tikzpicture}
        &
        \begin{tikzpicture}[scale = 1.78]
            \CropImage{\scene/\firstmethod}\secondcropxmin\secondcropymin\croplength{\secondcropcolor}
        \end{tikzpicture}
        &
        \begin{tikzpicture}[scale = 1.78]
            \CropImage{\scene/\firstmethod-MAPE}\secondcropxmin\secondcropymin\croplength{\secondcropcolor}
        \end{tikzpicture}
        &
        \RotTableMethodLabel{\firstmethoddisplaynamea}{\firstmethoddisplaynameb}
        \\
        % row 2
        \begin{tikzpicture}[scale = 1.78]
            \CropImage{\scene/\secondmethod}\firstcropxmin\firstcropymin\croplength{\firstcropcolor}
        \end{tikzpicture}
        &
        \begin{tikzpicture}[scale = 1.78]
          \CropImage{\scene/\secondmethod-MAPE}\firstcropxmin\firstcropymin\croplength{\firstcropcolor}
        \end{tikzpicture}
        &
        \begin{tikzpicture}[scale = 1.78]
            \CropImage{\scene/\secondmethod}\secondcropxmin\secondcropymin\croplength{\secondcropcolor}
        \end{tikzpicture}
        &
        \begin{tikzpicture}[scale = 1.78]
            \CropImage{\scene/\secondmethod-MAPE}\secondcropxmin\secondcropymin\croplength{\secondcropcolor}
        \end{tikzpicture}
        &
        \RotTableMethodLabel{\secondmethoddisplaynamea}{\secondmethoddisplaynameb}
        \\
        % row 3
        \begin{tikzpicture}[scale = 1.78]
            \CropImage{\scene/\thirdmethod}\firstcropxmin\firstcropymin\croplength{\firstcropcolor}
        \end{tikzpicture}
        &
        \begin{tikzpicture}[scale = 1.78]
            \CropImage{\scene/\thirdmethod-MAPE}\firstcropxmin\firstcropymin\croplength{\firstcropcolor}
        \end{tikzpicture}
        &
        \begin{tikzpicture}[scale = 1.78]
            \CropImage{\scene/\thirdmethod}\secondcropxmin\secondcropymin\croplength{\secondcropcolor}
        \end{tikzpicture}
        &
        \begin{tikzpicture}[scale = 1.78]
            \CropImage{\scene/\thirdmethod-MAPE}\secondcropxmin\secondcropymin\croplength{\secondcropcolor}
        \end{tikzpicture}
        &
        \RotTableMethodLabel{\thirdmethoddisplaynamea}{\thirdmethoddisplaynameb}
    \end{tabular}
    \end{minipage}
\end{tabularx}

    \caption{Equal-rendering-time comparison ($1/30$ seconds of computation) of \gls{pt} (left), \gls{mala} Restore (middle), and Diffusion Restore (right) for the \textsc{Swimming Pool} scene provided by \citet{rioux2020delayed}. The inset error maps show the corresponding \gls{mape} for each rendering.}
    \label{fig:pool}
\end{figure*}

\vphantom{\vspace{5cm}}

\appendix
\counterwithin{figure}{section}
\onecolumn

% ----------------------------------------------------------------------------------------------------
% additional results
% ----------------------------------------------------------------------------------------------------

\vphantom{\gls{pt}\gls{mala}}
\vphantom{\gls{mae}\gls{mse}\gls{mrse}\gls{mape}}

% ----------------------------------------------------------------------------------------------------
% Equal-rendering-time comparison
% ----------------------------------------------------------------------------------------------------
\section{Equal-rendering-time comparisons}\label{sec:equal-rendering-time}

\definecolor{myblue}{RGB}{0, 114, 178}
\definecolor{myorange}{RGB}{230, 159, 0}
\definecolor{myteal}{RGB}{0, 158, 115}
\definecolor{mypurple}{RGB}{204, 121, 167}

% \begin{figure}[!htbp]
    \includegraphics[width=\linewidth]{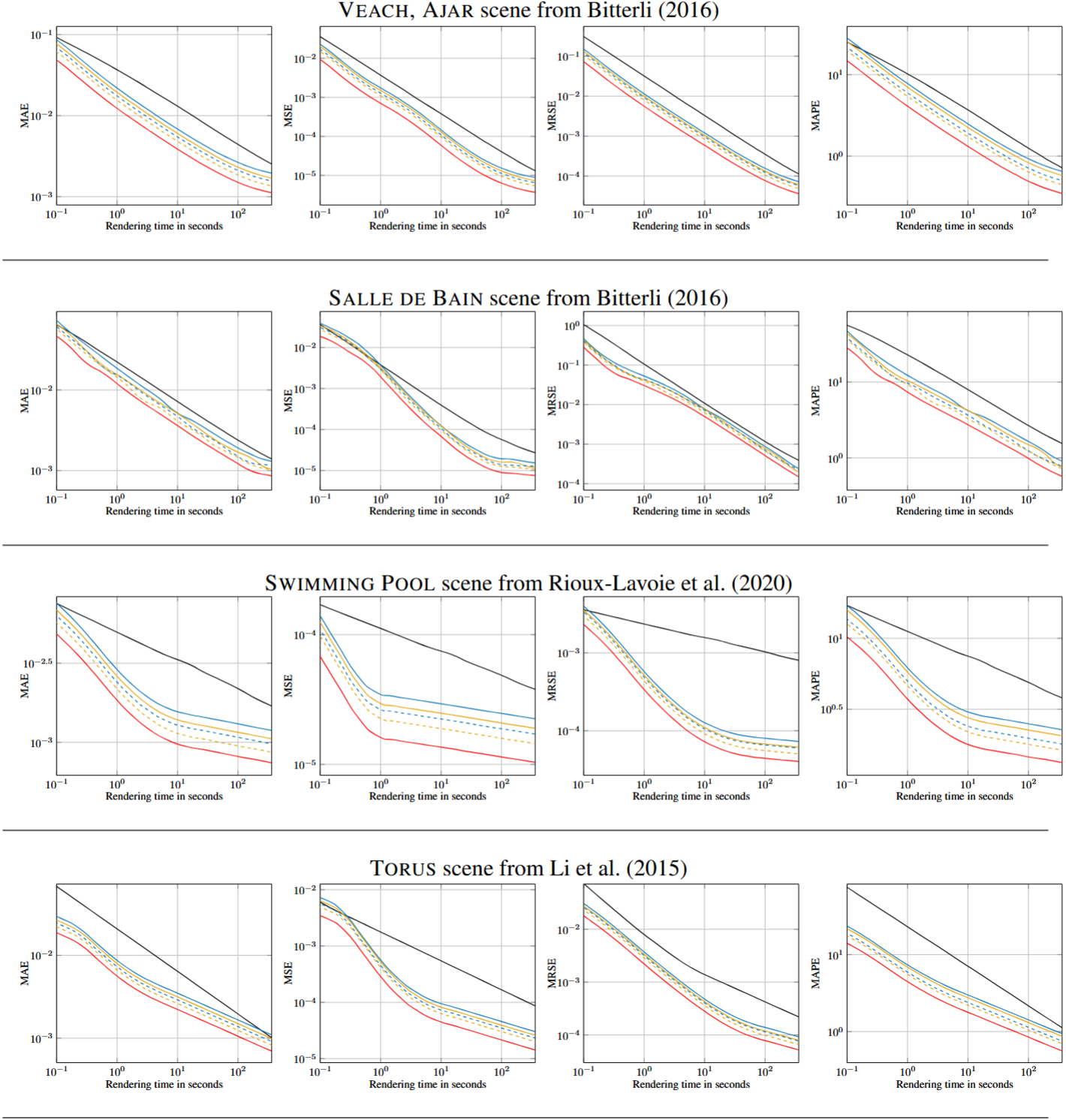}
    \clearpage
    \includegraphics[width=\linewidth]{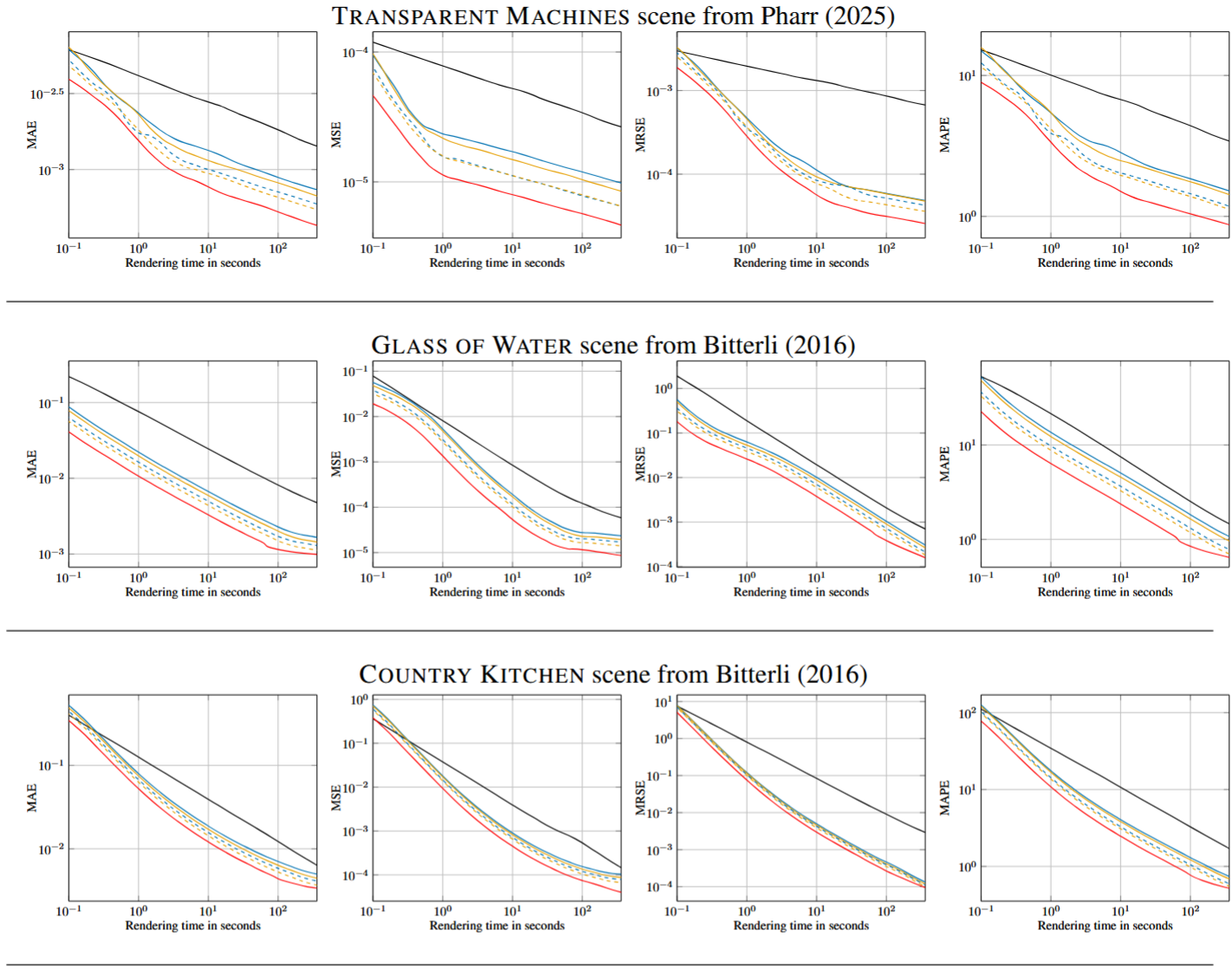}
    
    \captionof{figure}{Equal-rendering-time comparison (up to $360$ seconds, averaged over 100 realizations) across various scenes between \gls{pt}, Metropolis, \gls{mala}, Metropolis Restore, \gls{mala} Restore and Diffusion Restore. Plot points were sampled at logarithmically spaced rendering times of $10^{i/4}$ seconds.}
    \label{fig:equal-rendering-time}
% \end{figure}

\begin{table*}[h]
    \centering
    \renewcommand{\arraystretch}{1}
    \setlength{\tabcolsep}{3pt}
    % \begin{tabular}{lc | c | c | c | c | c | c | c}
    \begin{tabular}{lc | c | c | c | c | c | c}
        \toprule
        &\textsc{Veach, Ajar}
        &\textsc{Pool}
        &\textsc{Torus}
        &\textsc{Machines}
        % &\textsc{Water Caustic}
        &\textsc{Glass}
        &\textsc{Salle de Bain}
        &\textsc{Kitchen}\\
        \midrule
        \gls{pt}           & 0.05299261 		 & 0.08986768 		   & 0.57806207 		  & 0.04016843 		   & 0.81727877 		 & 0.43829467 		 & 0.31332345 		    \\
        Metropolis         & 0.01725706 		 & 0.02647645 		   & 0.25702271 		  & 0.04463808 		   & 0.91716730 		 & 0.27638444 		 & 0.44722915 		    \\
        Metropolis Restore & 0.00909867 		 & 0.02388700 		   & 0.05702231 		  & 0.007086883 	   & 0.34702540 		 & 0.12423548 		 & 0.15921618 		    \\
        \gls{mala}         & 0.01612867 		 & 0.02536953 		   & 0.22514550 		  & 0.01200780 		   & 0.87966343 		 & 0.23604773 		 & 0.40973712 		    \\
        \gls{mala} Restore & 0.00871587 		 & 0.02108429 		   & 0.04348912 		  & 0.00528136 		   & 0.29332845 		 & 0.08621837 		 & 0.15015922 		    \\
        Diffusion Restore  & \textbf{0.00727228} & \textbf{0.01998694} & \textbf{0.03971335} & \textbf{0.00475199} & \textbf{0.23602912} & \textbf{0.07803152} & \textbf{0.13811830}\\
        \bottomrule
    \end{tabular}
    \caption{Equal-rendering-time ($1/30$ second, averaged over 100 realizations) comparison of \gls{pt}, Metropolis, \gls{mala}, Metropolis Restore, \gls{mala} Restore and Diffusion Restore for the \textsc{Veach, Ajar} \citep{bitterli2016resources}, \textsc{(Swimming) Pool} \citep{rioux2020delayed}, \textsc{Torus} \citep{li2015anisotropic}, \textsc{(Transparent) Machines} \citep{pharr2025scenes},
    % \textsc{Water Caustic} \citep{bitterli2016resources},
    \textsc{Glass (of Water)} \citep{bitterli2016resources}, \textsc{Salle de Bain} \citep{bitterli2016resources} and \textsc{(Country) Kitchen} \citep{bitterli2016resources} scene.}
	\label{tab:equal-rendering-time}
\end{table*}

\newpage

% ----------------------------------------------------------------------------------------------------
% Equal-sample-count comparison
% ----------------------------------------------------------------------------------------------------
\section{Equal-sample-count comparisons}\label{sec:equal-sample-count}

% \begin{figure}[!htbp]
    \includegraphics[width=\linewidth]{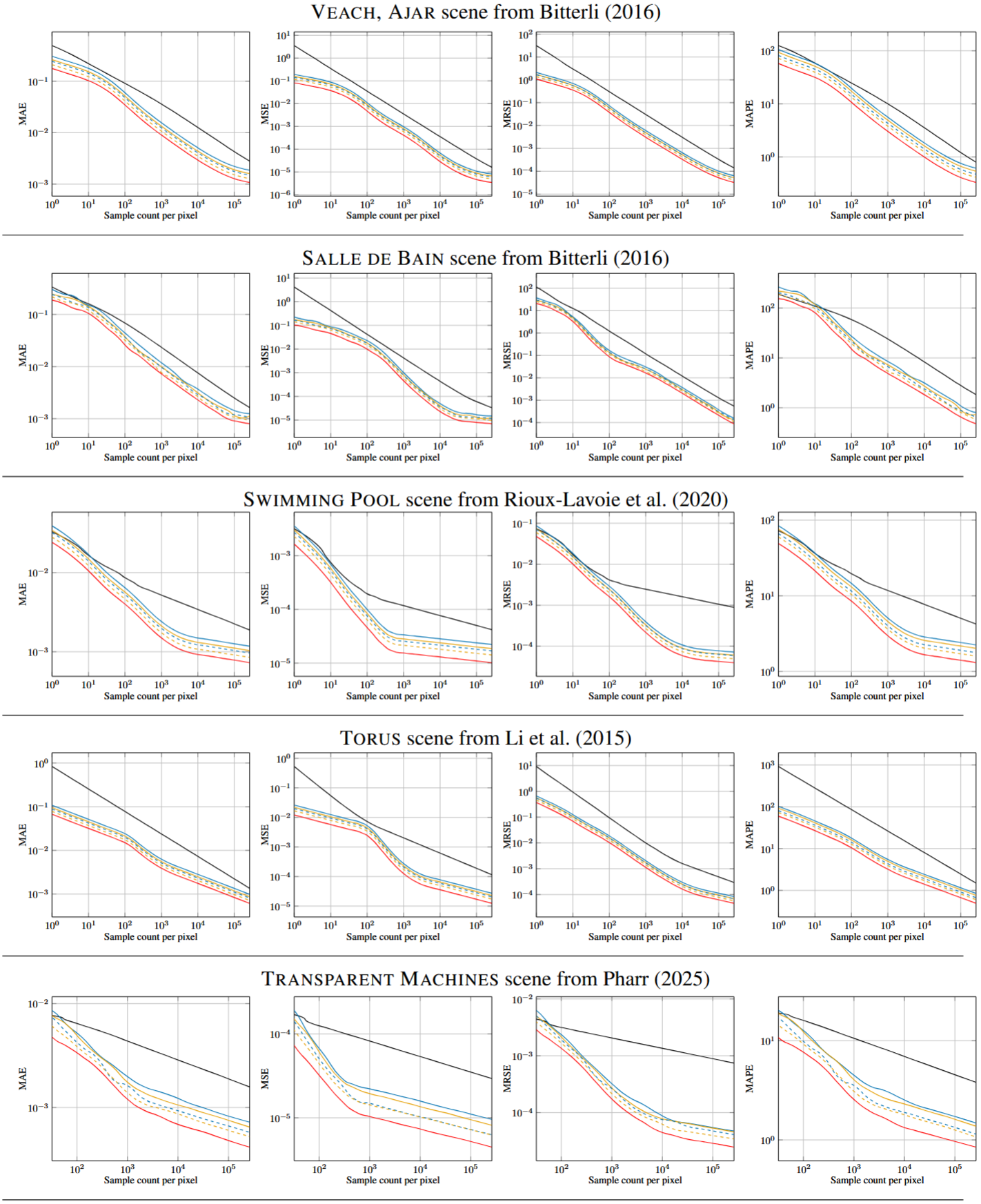}
    \clearpage
    \includegraphics[width=\linewidth]{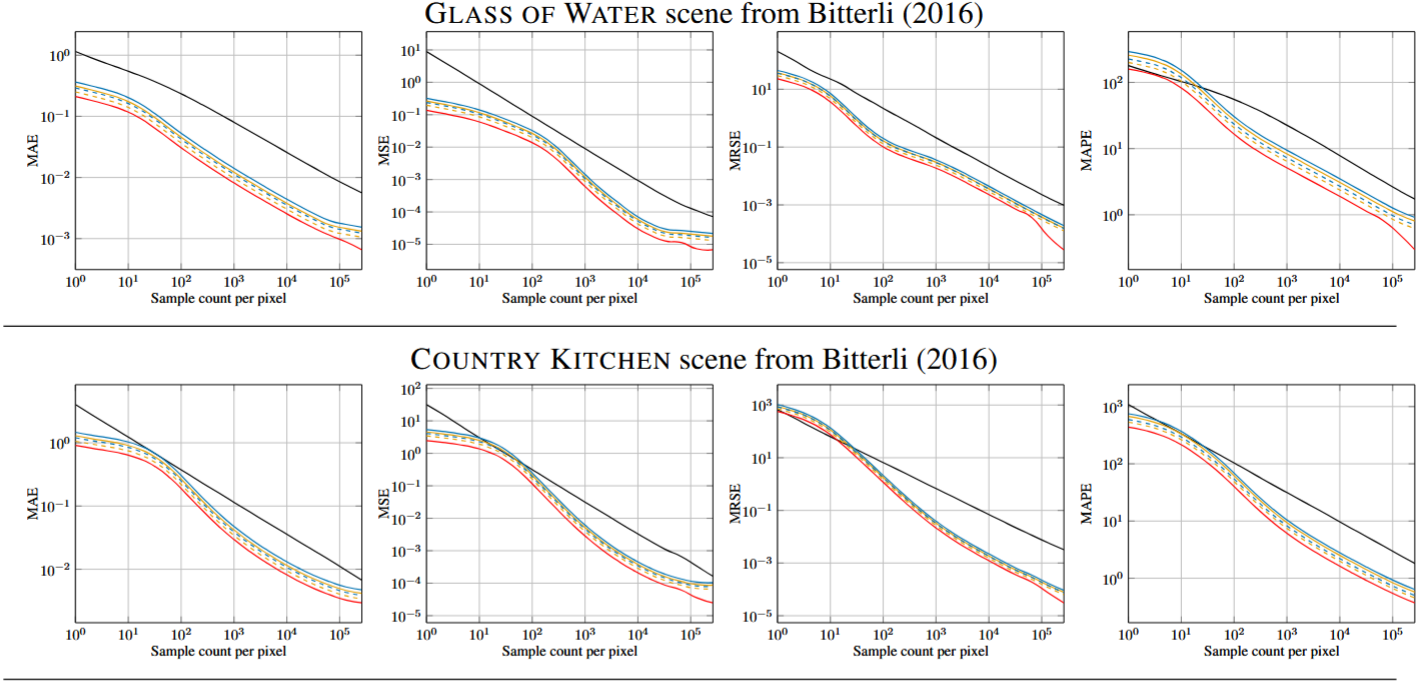}

    \captionof{figure}{Equal-sample-count comparison (up to $2^{18}$ \gls{spp}, averaged over 100 realizations) across various scenes between \gls{pt}, Metropolis, \gls{mala}, Metropolis Restore, \gls{mala} Restore and Diffusion Restore. Plot points were sampled at logarithmically spaced sample counts of $2^i$ \gls{spp}.}
    \label{fig:equal-sample-count}
% \end{figure}

\begin{table*}[h]
    \centering
    \renewcommand{\arraystretch}{1}
    \setlength{\tabcolsep}{4pt}
    % \begin{tabular}{lc | c | c | c | c | c | c | c}
    \begin{tabular}{lc | c | c | c | c | c | c}
        \toprule
        &\textsc{Veach, Ajar}
        &\textsc{Pool}
        &\textsc{Torus}
        &\textsc{Machines}
        % &\textsc{Water Caustic}
        &\textsc{Glass}
        &\textsc{Salle de Bain}
        &\textsc{Kitchen}\\
        \midrule
        \gls{pt}           & 0.00000939     	 & 0.00003696	       & 0.00008083 		 & 0.00002587 		   & 0.00005203 		 & 0.00002428 		   & 0.00008983		     \\
        Metropolis         & 0.00000823     	 & 0.00002219 		   & 0.00002712 		 & 0.00000957 		   & 0.00002170 		 & 0.00001477 		   & 0.00007410 		 \\
        Metropolis Restore & 0.00000608          & 0.00001489 		   & 0.00001757 		 & 0.00000629 		   & 0.00001204 		 & 0.00001152 		   & 0.00002907 		 \\
        \gls{mala}         & 0.00000674          & 0.00001863          & 0.00002274 		 & 0.00000816 		   & 0.00001780 		 & 0.00001274 		   & 0.00005741 		 \\
        \gls{mala} Restore & 0.00000498          & 0.00001117          & 0.00001326 		 & 0.00000625 		   & 0.00000915 		 & 0.00000967 		   & 0.00001139 	     \\
        Diffusion Restore  & \textbf{0.00000348} & \textbf{0.00000919} & \textbf{0.00000941} & \textbf{0.00000447} & \textbf{0.00000680} & \textbf{0.00000694} & \textbf{0.00000979} \\
        \bottomrule
    \end{tabular}
    \caption{Equal-sample-count ($2^{18}$ \gls{spp}, averaged over 100 realizations) comparison of \gls{pt}, Metropolis, \gls{mala}, Metropolis Restore, \gls{mala} Restore and Diffusion Restore for the \textsc{Veach, Ajar} \citep{bitterli2016resources}, \textsc{(Swimming) Pool} \citep{rioux2020delayed}, \textsc{Torus} \citep{li2015anisotropic}, \textsc{(Transparent) Machines} \citep{pharr2025scenes},
    % \textsc{Water Caustic} \citep{bitterli2016resources},
    \textsc{Glass (of Water)} \citep{bitterli2016resources}, \textsc{Salle de Bain} \citep{bitterli2016resources} and \textsc{(Country) Kitchen} \citep{bitterli2016resources} scene.}
    \label{tab:equal-sample-count}
\end{table*}

\newpage

% ----------------------------------------------------------------------------------------------------
% bias analysis
% ----------------------------------------------------------------------------------------------------
\section{Bias induced by the modified killing rate}

\begin{figure}[H]
    \centering
    \begin{subfigure}{.49\linewidth}
        \centering
        \includegraphics[width=\linewidth]{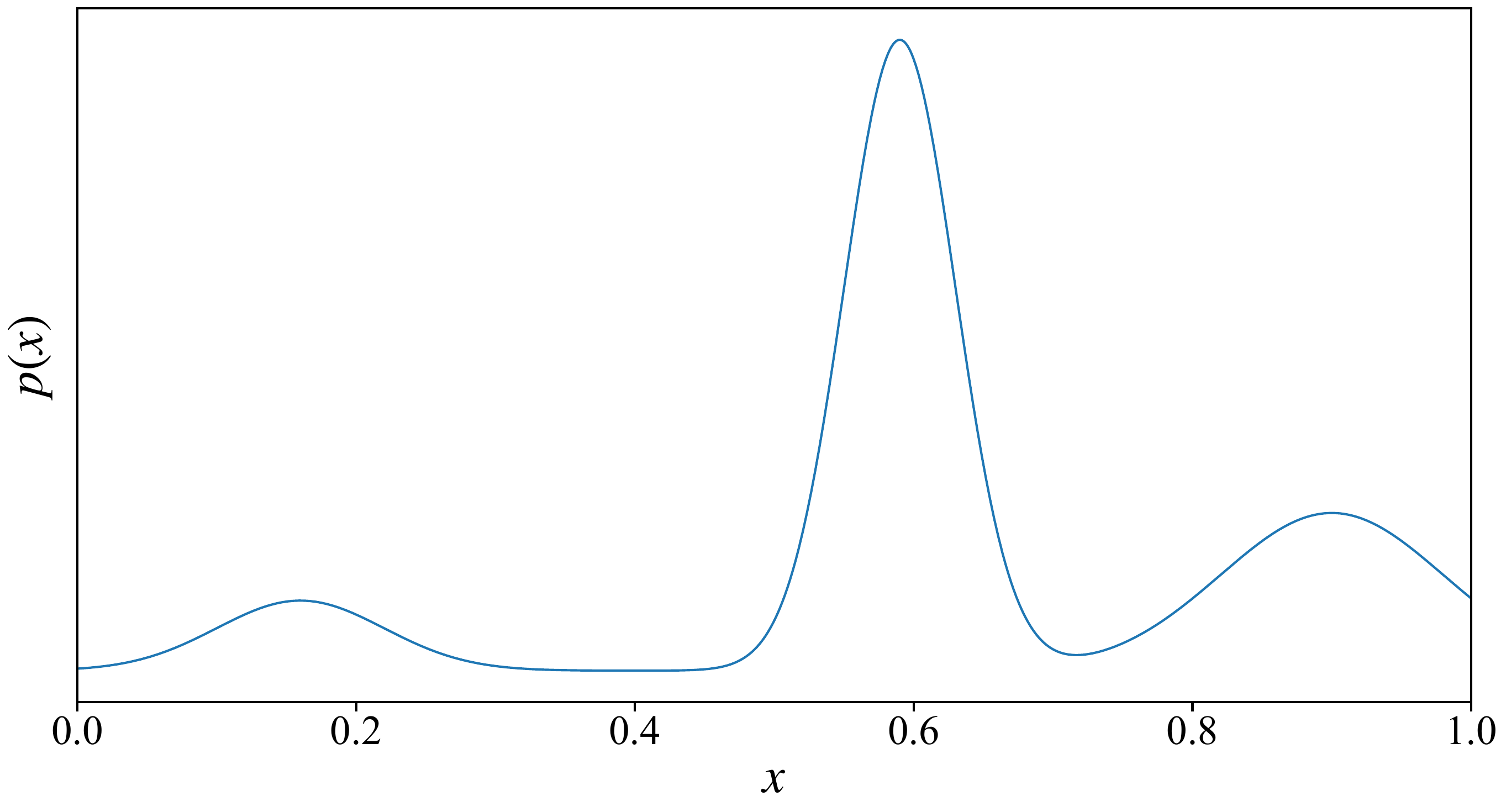}
    \end{subfigure}
    \begin{subfigure}{.49\linewidth}
        \centering
        \includegraphics[width=\linewidth]{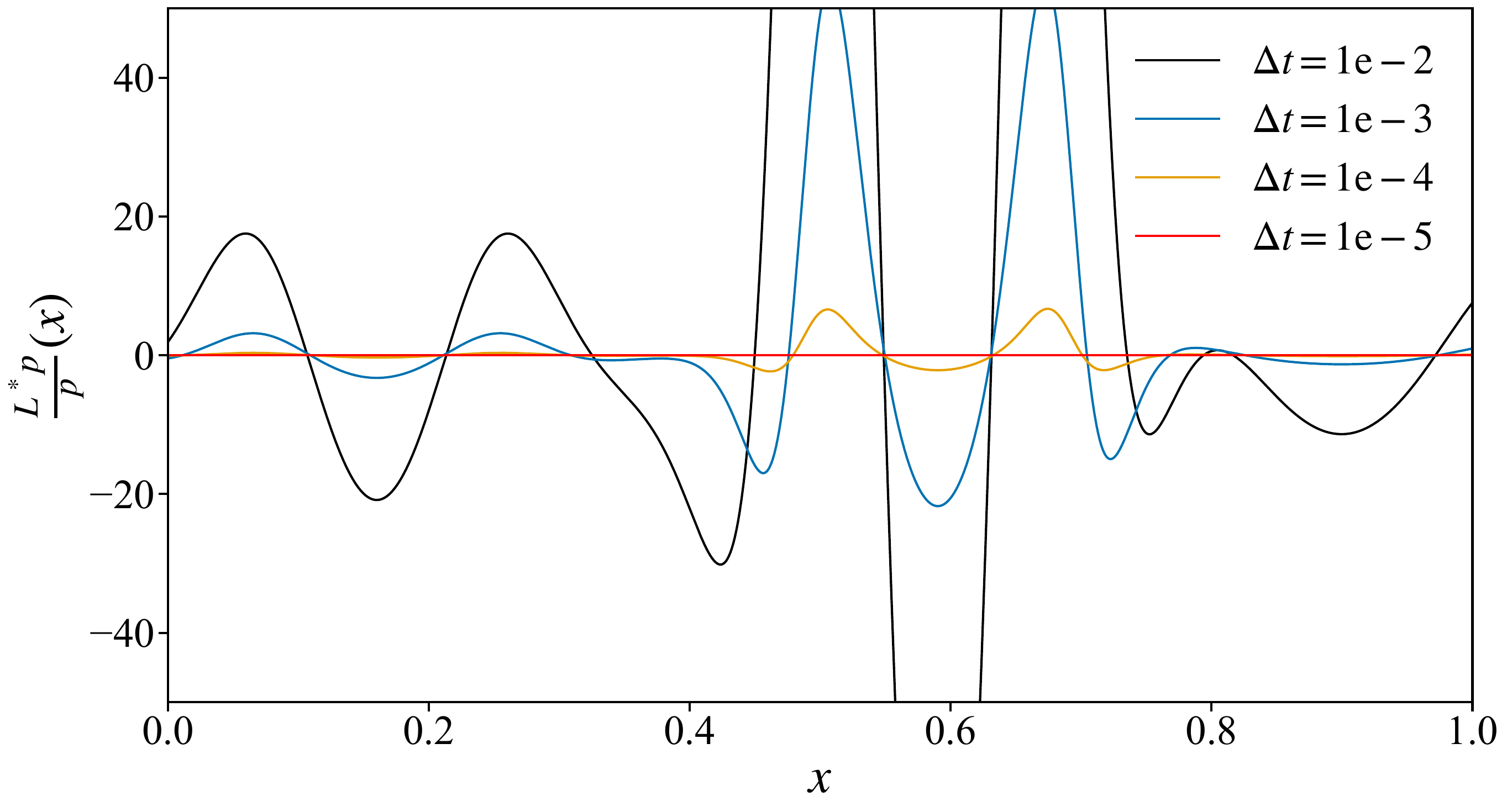}
    \end{subfigure}
    \caption{Left: Target density $\targetdensity$ exhibiting separated modes, as typically encountered in light transport settings. Right: Visualization of the bias term \eqref{eq:adjoint-bias}, $\localgenerator^\ast\targetdensity/\targetdensity$, for the local dynamics with generator $\localgenerator$ described in \Cref{sec:bias-analysis}, shown for different choices of $\Delta\timepoint$. In general, this term converges to zero as $\Delta\timepoint\to0+$, implying vanishing bias. As illustrated on the right, in practice the bias term is already negligible for sufficiently small $\Delta\timepoint$ (in this example, and throughout our numerical study in \Cref{sec:numerical-study}, $\Delta\timepoint=1\mathrm{e}{-}5$), effectively yielding zero bias.}
    \label{fig:bias}
\end{figure}

\end{document}